\documentclass[article]{jhep}

\usepackage{epsfig} 
\usepackage{amsmath}

\makeatletter
\def\lesssim{\mathrel{\mathpalette\vereq<}}
\def\vereq#1#2{\lower3pt\vbox{\baselineskip1.5pt \lineskip1.5pt
\ialign{$\m@th#1\hfill##\hfill$\crcr#2\crcr\sim\crcr}}}

\def\alt{\lesssim}

\makeatother

\conference{The Latin American Symposium on High Energy Physics}

\title{THE SUPERSYMMETRY SOFT-BREAKING LAGRANGIAN --- WHERE EXPERIMENT
 AND STRING THEORY MEET\footnote{Lectures at the Latin American School
SILAFAE III, April 2000, Cartagena, Colombia.}}

\author{G. L. Kane\\ Randall Laboratory of Physics 
University of Michigan, Ann Arbor, MI 48109\\
gkane@umich.edu}

\begin{abstract}
{\scriptsize After an introduction recalling that we expect low energy
supersymmetry to be part of our description of nature because of
considerable indirect evidence and successful predictions, and a
discussion of the essential role of data for formulating and testing
string theory, these lectures focus on the role of the supersymmetry
soft-breaking Lagrangian in connecting experiment and string theory.
How to measure tan$\beta$ and the soft parameters is examined via a
number of applications, and the difficulty of measuring tan$\beta$ at
hadron colliders is explained. In each case the important role of soft
phases is made explicit, and the true number of parameters is counted.
Applications include the chargino and neutralino sectors, the Higgs
sector and how its results change when phases are included, measuring
the (relative) gluino phase, CP violation in K and B systems and whether
all CP violation can be due to soft phases, how to learn if the LSP is
the cold dark matter of the universe, and baryogenesis.  It is
emphasized that the success of supersymmetry in explaining the breaking
of electroweak symmetry is probably its most important achievement, and
implications of that explanation for superpartner masses are shown.
Combining many of these considerations, a further application argues
that a lepton collider of total energy 600 GeV with a polarized beam is
one we can be confident will do important physics even after LHC.  The
question of the origins of CP violation, whether the CKM phase can be
zero, and the possibility that the soft phases can tell us about
compactification and supersymmetry breaking are discussed.  Some of the
applications and issues are examined in a D-brane based model that can
describe the usual collider and dark matter phenomenology, and includes
phases and CP violation as well.}
\end{abstract}
\newpage
\begin{document}

\hyphenation{SU-PER-SYM-ME-TRY SOFT-BREAKING LA-GRAN-GIAN}
\hyphenation{phys-i-cal re-pa-ra-me-ter-i-za-tion}
\hyphenation{LEP-TON COL-LID-ER}

\section{INTRODUCTION -- SUPERSYMMETRY? -- HOW CAN WE RELATE EXPERIMENT
AND STRING THEORY?}

We are confident the Standard Model of particle physics (SM) will be
extended, and its foundations strengthened, because

$\bullet$ it cannot explain the breaking of the electroweak symmetry
(EWSB) to accommodate mass

$\bullet$ it cannot explain the baryon asymmetry of the universe

$\bullet$ it cannot provide the cold dark matter of the universe

$\bullet$ it cannot incorporate gravity

$\bullet$ it cannot describe neutrino masses.

There are strong reasons to expect that low energy supersymmetry is
the probable outcome of experimental and theoretical progress, and
that it will soon be directly experimentally confirmed.  By low energy
supersymmetry, we mean (softly) broken supersymmetry with an effective
soft Lagrangian whose mass parameters are typically of order the
electroweak scale but otherwise not special or constrained.  It is
worthwhile to recall that the main reasons we take low energy
supersymmetry very seriously are not its elegance or its likely string
theoretical motivations, but its successful explanations and
predictions.  Confidence that low energy supersymmetry is part of the
correct description of nature is not a matter of belief; it is based on
evidence.  Of course, it could happen that all these successes are
coincidences, so we must either find superpartners and a light Higgs
boson, or demonstrate they do not exist(in which case low energy
supersymmetry does not describe nature).  The main successes of low
energy supersymmetry then are:

$\bullet$ Low energy supersymmetry, with string-motivated boundary
conditions (some quark Yukawa couplings such as the top Yukawa
coupling of order unity, and no $\mu$ term in the superpotential
because the theory describes massless strings), can provide the
explanation of how the electroweak symmetry is broken.  First, we write
the SM effective scalar potential as $V={m_\phi}^2\phi^{2} +\lambda
\phi ^{4}$, then supersymmetry first requires that $\lambda
=(g_{1}^{2}+g_{2}^{2})/2$, where $g_{1}$ is the U(1) coupling and
$g_{2}$ is the SU(2) coupling. Second, ${m_\phi}^2$ is not special at
the unification scale, but runs to be negative by the electroweak
scale, driven negative by the large top quark Yukawa coupling.  Thus
the ``Mexican hat'' potential with a minimum away from $\phi =0$ is
derived rather than assumed.  As is normal for progress in physics,
this explanation is not from first principles, but it is an explanation
in terms of the next deeper effective theory, and depends on the
supersymmetry breaking leading to soft masses of order the weak scale.
That is not circular because the crucial issue is how the
SU(2)$\times$U(1) symmetry is broken.   

$\bullet$ Supersymmetry can stabilize the hierarchy between the weak
scale and the unification scale, and mitigate the naturalness problem,
providing a small ratio of the weak scale to the unification scale,
because of log running.

$\bullet$ Supersymmetry allows the gauge couplings to unify.

$\bullet$ Supersymmetry can provide a cold dark matter candidate.

$\bullet$ Supersymmetry can explain the baryon asymmetry of the
universe. 

$\bullet$ Supersymmetry has made several correct predictions:

\hspace{.1in} (1) Supersymmetry predicted in the early 1980s that the top
quark would be heavy, because that was a necessary condition for the
validity of the EWSB explanation.

\hspace{.1in} (2) Supersymmetry correctly predicted the present experimental
value of $\sin ^{2}\theta _{W}$ before it was measured.

\hspace{.1in} (3) Supersymmetry requires a light Higgs boson to exist, and
the precision measurements lead to $M_{h}\alt 200$ GeV.

\hspace{.1in} (4) When LEP began to run in 1989 it was possible to say
that either LEP would discover light superpartners if they were very
light, or because all supersymmetry effects at LEP are loop effects, and
supersymmetry effects decouple as superpartners get heavier, there would
be no significant deviations from the SM discovered at LEP.  That is, it
is only possible to have loop effects large enough to measure at LEP +
SLC if superpartners are light enough to directly observe.  In
non-supersymmetric approaches it was natural to expect effects at LEP.

It is remarkable that supersymmetry was not invented to explain any of
the above physics.  Supersymmetry was written as an interesting and
beautiful property of theories, studied for its own sake, in the early
1970s.  Only after several years of improving the understanding of the
theory did people begin to realize that supersymmetry solved the above
problems, one by one.  It is also remarkable that all of the above
successes can be achieved simultaneously, with one consistent form of
the theory and its parameters.  Finally, it is also noteworthy that low
energy softly broken supersymmetry has no known incorrect predictions;
it is not so easy to write a theory that explains and predicts some
phenomena and has no conflict with other phenomena.

Some physicists argue that gauge coupling unification is the most
important success of supersymmetry, and it is indeed a major result.
But the issue of how to break the electroweak symmetry is the more
fundamental problem.  Any solution of that problem must simultaneously
accommodate both the boson and fermion masses.  Explaining the mechanism
of EWSB is the deepest reason why we should expect low energy
supersymmetry in nature.  No other approach should be taken to be of
comparable interest to supersymmetry unless it can provide an
appropriate explanation of EWSB.

It should already be clear that our framework for these lectures is the
traditional one with the Planck scale above $10^{18}$ GeV, and gauge
coupling unification somewhat above $10^{16}$ GeV.  We are convinced
that a consistent quantum theory of gravity and the SM forces requires
extra dimensions in some sense, but not that any of them are larger than
the (inverse) unification scale.  We expect the extra dimensions to be
small because the traditional picture based on having a primary theory
at the traditional Planck scale, with a hierarchy protected by
supersymmetry, provides beautiful, understandable explanations (such as
explaining EW symmetry breaking) and successful predictions, while
alternative approaches do not.  Alternative approaches should be
explored, but this approach should remain the default.

String theory can provide a consistent quantum theory of gravity.  In
addition it can incorporate the known forces and fundamental particles,
providing a corresponding theoretical framework.  But, string theory
will be formulated with extra space-time dimensions.  It must be
compactified to 4D, and supersymmetry must be broken, to give an
effective theory at the unification scale.  It is that effective theory
that is the focus of these lectures.

The region between the weak or collider scale, and the unification
scale, need not be empty --- indeed, we expect a variety of states in
that region, including right-handed neutrinos involved in generating
neutrino masses, possible axions, possible vector or SU(5) multiplets,
and more.  All that is assumed is that the theory remains perturbative
in the region from a TeV or so to the unification scale.  There is
strong evidence for this assumption --- both the unification of the
gauge couplings and the explanation of EWSB independently imply that the
theory is indeed perturbative in this region.

With this worldview we can divide the physics opportunities into three
categories.  The foundation is experiments, of course, experiments at
colliders such as LEP and the Tevatron and LHC, and b-factories, and
rare decays, and proton decay, and electric dipole moment
measurements, and cold dark matter detectors.  We will discuss all of
these in the following.  At the other end is the effort to construct a
consistent string theory which describes our world, which includes
finding out how it is compactified, and how supersymmetry is broken.  We
will see that there is a mismatch between these two efforts.  String
theorists and experimenters cannot communicate with each other, but
they can communicate through  the third category, supersymmetry
theorists.  The stheorists are needed for two reasons.  Experimenters
will measure, once superpartners and Higgs bosons are directly
observed, some kinematical masses of superpartner mass eigenstates,
and some cross sections times branching ratios. Essentially none of
the quantities that experimenters measure can emerge from a string
theory (as we will see more clearly below).  The stheorists will first
help convert the observables into the parameters of the supersymmetry
Lagrangian, and then will help convert that Lagrangian from its value
at the weak scale to its value at the unification scale.  That can in
turn be compared with the predictions of a compactified string theory
with broken supersymmetry.  Of course, a given person can contribute
in more than one category --- for example, experimenters can translate
the observables into the Lagrangian parameters, perhaps string
theorists can help formulate how to relate the weak scale Lagrangian
to the unification scale one, and stheorists can help formulate string
based models.  

In practice, however, I will argue that it is highly
probable that the historical path will be followed. String theory in
10D has no known testable predictions or explanations of particle
physics phenomena in the 4D world.  Probably figuring out how to
compactify and how to break supersymmetry is too difficult for anyone to
do purely from first principles.  But even more, suppose someone had
figured it out.  How would they convince themselves or anyone else
they had done so?  Many compactifications or brane worlds have three
families, so that is not a unique clue.  Most approaches to solving
the cosmological constant problem are generic.  Calculating the small
Yukawa couplings is likely to depend on many difficult details of the
theory, as is calculating the mass of the LSP or tan$\beta$.  I am not
arguing that string theory is not testable physics, or that we will
not be able to make great progress toward becoming confident that
string theory is the right approach to discovering the primary theory,
but that the crucial steps of learning why 3 space dimensions are
large and how supersymmetry is broken will almost certainly only come
after there is data on superpartners to point the way.  

In fact, the main result that will emerge from any tentative
compactification and supersymmetry breaking is the soft-breaking
supersymmetry Lagrangian, ${\cal L}_{soft},$ with its many parameters.
But until ${\cal L}_{soft}$ is measured it will not be possible to
recognize its correctness.  I expect it will be the other way around.
Once ${\cal L}_{soft}$ is measured and translated to the unification
scale (a significant challenge), its patterns will lead string theorists
to recognize how to compactify and how to break supersymmetry and how to
find the correct vacuum.  The data and the stheorists will be essential
to make progress.  At the same time, it will be necessary to have
understood and developed string theory itself to have the technology to
join onto the data and to recognize the significance of the patterns
that emerge from the data.  In the following we will look briefly at
some of the kinds of patterns and what they might imply.  Thus these
lectures focus on ${\cal L}_{soft}.$

\section{THE SUPERSYMMETRY SOFT-BREAKING LAGRANGIAN, ${\cal L}_{soft}$
-- LARGE PHASES?}

We know that supersymmetry is a broken symmetry, but the physics of
supersymmetry breaking is not yet understood.  That's not surprising ---
the symmetry breaking was the last thing understood for the Standard
Model (SM) too (assuming we do indeed understand it).  For the SM the
symmetry breaking originates in a different sector, with the
soft-breaking supersymmetry terms that make up the SM effective scalar
potential.  The symmetry breaking is transmitted to the SM bosons and
fermions by interactions with the Higgs field.  There were several ways
the EW symmetry could be broken --- we will not know for sure which is
nature's choice until a light Higgs boson is directly observed.  In the
supersymmetry case there are also several ways the supersymmetry could
be broken --- we do not know which, if any, is correct, and we do not
yet know how the breaking is transmitted to the superpartners.  All of
this information, both the breaking and the transmission, is encoded in
${\cal L}_{soft}$.
 
Fortunately, the theory is powerful enough that it is possible to write
the ``most'' general effective Lagrangian $\cite{S. Dimo}$ even though
we do not know the mechanism of supersymmetry breaking.  This Lagrangian
is defined to include all allowed terms that do not introduce quadratic
divergences in the theory, which include all allowed terms of dimension
two and three.  It is gauge invariant and Lorentz invariant.  The
Lagrangian depends of course on what the gauge group and particle
content are assumed to be.  We will study the case of the general
supersymmetric Standard Model (SSM), where we assume the theory contains
the SM particles and has the SM SU(3)$\times$SU(2)$\times$U(1) gauge
group, has two Higgs doublet fields as is necessary in supersymmetry to
give mass to both up and down type quarks, and has a conserved R-parity.
``Most'' is in quotes above for two reasons.  First, the correct theory
could be larger than the SSM, and second, some kinds of terms are not
normally included since they are usually very small in models
$\cite{D.R.T.}$.  If we wanted to extend the theory, for example to add
an extra singlet scalar or an additional U(1) symmetry, we could add the
associated terms.  Similarly, just as it is necessary to add new fields
such as right-handed neutrinos to the SM to incorporate neutrino masses
in the SM, we would have to add such fields and their superpartners and
the associated terms in ${\cal L}_{soft}$ if we wanted to include
neutrino masses in our theory.  We will not do that in these lectures.

The situation is analogous to that in the SM with the quark masses and
the CKM matrix which contains flavor mixing angles and a phase.  In the
supersymmetric case there are parameters that are masses, flavor
rotation angles, and phases.  Just as for the CKM matrix, all these
parameters have to be measured, unless a compelling theory determines
them eventually.  Before the top quark mass was known, we could assume a
value for the top quark mass and then calculate its production cross
section, its decay branching ratios and signatures, and all aspects of
its behavior.  Since all other needed SM parameters were measured only
the top mass was unknown; if some other SM parameters had not yet been
measured we would have had to choose various values for them too.  The
situation for superpartners is similar --- for any given set of masses
and flavor mixing angles and phases we can calculate the observable
aspects of superpartner behavior.  We can study any tentative
supersymmetry signal and decide if it is consistent with the theory, we
can make predictions and plan for future facilities, and much more.
${\cal L}_{soft}$ will contain at least 105 new parameters, depending on
what we include.  That might seem like a lot, but almost all have clear
physical interpretations.  Once there is data most will be measured, and
their patterns will quickly reduce the number.  The historical situation
with the SM was similar.  Once we knew the effective Lagrangian was
$V-A$ many parameters disappeared, and such a structure led to
recognizing it was a gauge theory which reduced the number more.
Probably the situation will be similar for supersymmetry.

[Counting parameters depends on assumptions.  One reasonable way to
count the SM parameters for comparison with supersymmetry is to assume
that all the particles are known, but not their masses or interactions.
Then the W and Z vertices can each have a space-time tensor character of
scalar, vector, etc (S, V, T, A, P), and each can be complex (so
multiply by 2).  Conserving electric charge, the Z can have 12 different
flavor-non-changing vertices for the 12 quarks and leptons (e,$\mu,
\tau, \nu_e, \nu_{\mu}, \nu_{\tau}, u, c, t, d, s, b$), plus 12
additional flavor-changing vertices ($e\mu, e\tau, \mu \tau,$ etc.).
This gives 240 parameters (12x5x2x2).  Similar counting for the W gives
180.  There are 12 masses.  Self-couplings of W and Z allowing CP
violation give 10.  The total here is 442 parameters.]

We can write the soft Lagrangian as

$$-{\cal L}_{soft}=\frac{1}{2}(M_{3}\tilde{g}\tilde{g}+M_{2}\tilde{W}
\tilde{W}+M_{1}\tilde{B}\tilde{B}+h.c.)$$
$$+\tilde{Q}^{\dagger}M_{\tilde{Q}}^{2}\tilde{Q}
+{\tilde U}^{c\dagger}M_{\tilde U}^{2}\tilde U^{c}
+{\tilde{D}}^{c\dagger}M_{\tilde{D}}^{2}\tilde{D}^{c}
+{\tilde{L}}^{\dagger}M_{\tilde{L}}^{2}\tilde{L}
+\tilde{E}^{c\dagger}$$
$$M_{\tilde{E}}^{2}\tilde{E}^{c}+\tilde
+U^{c}\tilde{A}_{U}\tilde{Q}H_{U}+\tilde{D}^{c}\tilde{A}_{D}
\tilde{Q}H_{D}+\tilde{E}^{c}\tilde{A}_{E}\tilde{L}H_{D}$$
\begin{eqnarray}
+h.c.+m_{H_{U}}^{2}H_{U}^{\ast}H_{U}+m_{H_{D}}^{2}H_{D}^{\ast}H_{D}
+(bH_{U}H_{D}+h.c.).
\end{eqnarray}

Supersymmetry is broken because these terms contribute explicitly to
masses and interactions of (say) winos or squarks but not to their
superpartners.  The underlying supersymmetry breaking is assumed to be
spontaneous, but its effects are explicit in this effective Lagrangian.
Similarly, how supersymmetry breaking is transmitted to the
superpartners and their interactions is encoded in the parameters of
${\cal L}_{soft}$.  All the quantities in $ {\cal L}_{soft}$ are scale
dependent and satisfy known renormalization group equations.  ${\cal
L}_{soft}$ has the same form at any scale.  All of the coefficients in
${\cal L}_{soft}$ --- the ``soft'' parameters --- can be complex, as
long as the Lagrangian is Hermitian.

The gaugino masses $M_{i}$ can be written as $M_{3}=\left| M_{3}\right|
e^{i\phi _{3}}$, etc.  From now on in these lectures we will for simplicity
write $M_{3}$ for the magnitude and assume the phase is explicitly present.

The squark and slepton mass matrices of Eq.(2.1) are Hermitian matrices,
so the diagonal elements are real, while the off-diagonal elements can
be complex.  The absence of flavor changing neutral decays of quarks and
leptons puts strong constraints on these off-diagonal elements
$\cite{For a r}$.  The trilinear or triscalar couplings
$\tilde{A}_{U,D,E}$ are general 3x3 complex matrices in flavor space.

Symmetries of the theory allow some of the parameters to be absorbed or
rotated away with field redefinitions.  The net number left
$\cite{D. Sutter}$ is 103.  If the gravitino is the lightest
superpartner (LSP) then its mass and phase have to be included.
Finally, although there is no bare $\mu $ term in the superpotential in
a string-based approach, because the terms present are associated with
massless strings, it is both necessary phenomenologically and expected
theoretically that an effective $ \mu $ term will be generated, with the
magnitude of $\mu $ of a size similar to the other soft terms.  If there
were no effective $\mu$ term, the Higgs sector would have a massless
pseudoscalar that does not exist.  In general there will also be a phase
$\phi _{\mu}$.

Finally, then, we will examine the behavior of phenomena based on
supersymmetry physics that results from a superpotential with Yukawa
coupling and an effective $\mu$ term,

\begin{eqnarray}
W=y_{U}QU^{c}H_{U}+y_{D}QD^{c}H_{D}
+y_{E}LE^{c}H_{D}+\mu_{eff}H_{u}H_{d}
\end{eqnarray}

plus a soft Lagrangian ${\cal L}_{soft}$ (from now on we write
 $\mu_{eff}=\mu$).

We will focus on the phases of the soft parameters in these lectures,
for several reasons. They are the least studied aspects of the soft
parameters.  (The off-diagonal flavor structure of the squark and
slepton soft masses and of the triscalar terms is also little-studied
and quite interesting.)  If the phases are large they can have very
large effects $\cite{M. Brhlik}$ on a variety of interesting phenomena
--- they generate CP violation, they affect the baryon asymmetry of the
universe, the relic density and detectability of cold dark matter, rare
decays, implications of the Higgs sector, and superpartner masses, cross
sections, and branching ratios.  The patterns of the phases, and whether
they are measured to be large or small, will point to compactifications
on different manifolds and to different ways to break supersymmetry.  We
now understand how to measure most of the phases.  In keeping with the
view that the phases must first be measured before their implications
for compactification and supersymmetry breaking can be understood, we
will emphasize in the following how to measure the phases and how they
affect observables. For a very large variety of particle physics and
cosmology phenomena one can be badly misled if phases are large but are
not included in analyses.

\section{HOW DO THE SOFT PARAMETERS SHOW UP IN EXPERIMENTS?}

Let's suppose superpartners and Higgs bosons are found.  First let's
list the masses of the particles that will eventually be seen.  There
are 4 neutralino masses, associated with the soft terms from
$W^{0},B^{0},H_{U}^{0},H_{D}^{0}$ (or, in the mass eigenstate basis,
$\gamma ,Z,H_{U}^{0},H_{D}^{0}).$ The neutralino superpartners mix, so
the physical neutralino mass eigenstates $ \tilde{N}_{1,2,3,4}$ will be
the eigenvalues of a mass matrix.  We'll write this matrix later.
Similarly, there are 2 chargino mass eigenstates from the chargino mass
matrix.  There are 4 Higgs boson masses, for $ h^{0},H^{0},A^{0},H^{\pm
}.$ There is 1 gluino mass, and 1 gravitino mass.  The squark mass
matrix for up-type squarks has 6 independent eigenvalues, the
superpartners of the left- and right-handed quarks u,c,t: $\tilde
{u}_{L},\tilde{c}_{L},\tilde{t}_{L},\tilde{u}_{R},\tilde{c}_{R},
\tilde{t}_{R}.$ Similarly, there are 6 down-type mass eigenstates and 6
charged lepton mass eigenstates.  In the SSM there are only the three
left-handed neutrinos and their sneutrinos.  These add up to 33 physical
masses that can be measured if all the states are found in experiments.
A 34th mass is the gravitino mass; if the gravitino is not the LSP then
it may not be possible to measure its mass since it couples too weakly
to be produced directly at colliders, and affects only some
early-universe cosmology.

Another parameter is $\tan \beta $, the ratio of the vacuum expectation
values (vev's) of the two Higgs fields,

\begin{eqnarray}
\hspace{.25in}\tan \beta =\left\langle H_{U}\right\rangle /\left\langle
H_{D}\right\rangle .
\end{eqnarray}

$\tan \beta $ is intrinsically a low energy parameter, since the Higgs
fields do not have vev's until the RGE running induces them somewhat
above the weak scale.  We will see that measuring $\tan \beta $ is in
general difficult, and cannot be done accurately without a lepton
collider with a polarized beam that is above the threshold for several
superpartners.  When we are trying to deduce the unification scale
Lagrangian $\tan \beta $ can be traded for a high scale parameter in the
Higgs sector.

43 of the parameters in ${\cal L}_{soft}$ are phases.  As we  have
mentioned, they affect essentially all observables. 
Knowledge of the phases could provide the essential clues to how to
compactify and break supersymmetry.

Phenomenologically, life would be much simpler if the phases were zero,
or small.  It would be much easier to determine the soft parameters from
data, to measure $\tan \beta $, etc.  Are there reasons to think the
phases might be small?  Until recently $\cite{T. Ibrahim,G. Good}$ it
was thought that the absence of signals for electric dipole moments
(EDMs) of neutrons and electrons implied that some flavor diagonal
phases were small $\cite{M. Dugan}$.  These arguments concerning EDM
phases would not impact many of the phases even if they did apply.
Basically, if all phases and soft parameters were considered as
independent and if first-family squarks and sleptons were not heavier
than a few hundred GeV, then large phases would imply EDMs larger than
the experimental limits.  Of course the first family squarks and
sleptons could be heavy, suppressing EDM's, but let us assume they are
not.  This issue has been reexamined
$\cite{T. Ibrahim,G. Good,S. Pokorski,L. Everett}$.  In a real theory it
would necessarily occur that various soft terms and phases are related,
perhaps originating from a few F-term vev's or some other mechanism.
Some cancellations are automatic but were not included in the original
estimates, such as those between charginos and neutralinos arising from
the automatic negative sign in the Lagrangian between the
${H_{U}^{+}}{H_{D}^{-}}-{H_{U}}^{0}{H_{D}}^{0}$ terms.  Others have been
shown to arise in models (such as the D-brane model $\cite{L. Everett}$
presented later in the lectures).  Some people have worried that the
needed cancellations are a fine-tuning.  That is not so --- any theory
of the soft parameters will give relations among them.  What would
happen here if such cancellations indeed occurred is somewhat analogous
to the GIM mechanism.  If such cancellations occur, it will not be
because of an obvious symmetry, but rather a symmetry at the string
scale that leads to relations among the soft parameters.  The bottom
line is that the flavor-diagonal phases could well be large and still be
consistent with current EDMs.  However, if the EDM experiments improve
approximately an order of magnitude and still do not see a signal then
the phases they constrain are probably small.

At present there are also reasons why it seems that the phases may be
large.  In general this has to do with how CP violation effects originate.

\hspace{.1in} A. The baryon asymmetry of the universe cannot be
explained in the SM $\cite{M. B. Gavela}$, so some other significant
phases are needed.  The role of CP violation in various ways of
generating a baryon asymmetry is not yet well enough understood to be
sure of the origin of the phases.  The supersymmetry soft phases provide
the CP violation if the baryon asymmetry is generated at the electroweak
phase transition, and perhaps also if the baryon asymmetry is generated
by an Affleck-Dine mechanism.  If the baryon asymmetry is generated at
the EW phase transition, the size of the needed phases depends on how
much washout occurs, and on theoretical issues that are not yet
settled. Large phases may be required.  In the case of decay of heavy
Majorana neutrinos there is no concrete model that allows one to discuss
the origin of the CP violation definitively, and it could be from soft
phases or from lepton Yukawas.  Explaining the baryon asymmetry is
certainly easier with large soft phases, and may require them.

\hspace{.1in} B. Normally it is assumed that the SM CKM phase accounts
for CP violation in the Kaon system, which requires a large CKM phase,
and the b-factories will study whether the CKM phase describes CP
violation in the B system.  If the CKM phase turns out to be small
$\cite{J.-M. Frere}$, supersymmetry soft phases are then clearly needed
to describe CP violation in the Kaon and B systems (they can do that,
see below).  Since the Yukawa couplings enter the superpotential, and
are present even if supersymmetry is unbroken, it is easy to imagine
that the Yukawa couplings are relatively real so that the CKM phase is
zero (it may get small renormalization effects at the weak scale but
would still be effectively zero numerically).  The soft phases can enter
as a result of the supersymmetry breaking, which can be transmitted to
the observable sector by complex vev's of F-terms, analogous to complex
Higgs vev's, giving complex soft terms but relatively real Yukawas.

\hspace{.1in} C. If the CKM phase is large, more data is needed to see
if there is evidence for soft phases in the Kaon and B systems.  But if
the CKM phase is large and one looks at how the soft phases are
calculated from the superpotential (and Kahler potential and gauge
kinetic function), it is easy to see how the soft phases could be large
$\cite{For a different}$.  In general, the Yukawas are functions of
moduli that get possibly complex vev's, and soft triscalar terms and
gaugino masses are functions of those moduli vev's and also F-term vev's
that can be complex.  In particular, the triscalar $\tilde{A}$ terms are
linear combinations of Yukawas and derivatives of Yukawas, so if the CKM
phase is large it is likely the $ \tilde{A}$ phases are large (care must
be taken here to be precise about whether one is discussing the
$\tilde{A}$ of Eq.(3.1) or the coefficients $A_{ijk}$ if one writes
$\tilde{A}_{ijk}=A_{ijk}Y_{ijk}$).  If the vev's that give rise to the
spontaneous supersymmetry breaking are complex, it is very likely some
soft phases will be large.

The form for ${\cal L}_{soft}$ is rather general, and allows for some
other possibly important effects, such as D-terms (from the breaking of
extra U(1) symmetries) that give contributions to squark and slepton
masses, or Planck scale operators that lead to contributions to masses
when some fields get vev's.  Extra U(1)'s or extra scalars can lead to a
larger neutralino mass matrix than the 4x4 one expected here.  Terms of
the form $\phi ^{\ast }\phi ^{2}$ (rather than $\phi ^{3}$) are
generally allowed $\cite{D.R.T.}$ in gauge theories where the scalars
are charged under some gauge group, but no models are yet known where
such terms give significant effects.  They can be added if necessary.

I have said that it is very unlikely that string theorists could guess
the correct compactification and supersymmetry breaking mechanism and
convince themselves or anyone else that they had the answer.  There is
an additional difficulty with a top-down approach.  To predict any weak
scale observable it is necessary to know several soft parameters, as we
will see in detail below.  In addition, to run down to the weak scale it
is usually necessary to know several more.  Further, the $\mu $
parameter is likely to have a different physical origin (e.g. from the
Kahler potential as in the Giudice-Masiero mechanism) and it also has to
be known.  The phases have to be known.  Finally, it is necessary to
calculate $\tan \beta $ which requires additional soft parameters and
their running.  Basically, it is necessary to get it all right at once
to predict any weak scale observable correctly.  Conversely, if any part
of the assumed compactification and supersymmetry breaking approach is
wrong, predictions for some or all observables are likely to be wrong in
practice.  This reinforces the view that it is likely that it will be
necessary to measure the parameters of $ {\cal L}_{soft}$ at the weak
scale in order to formulate string theory in 4D with broken
supersymmetry.

\section{MEASURING THE SOFT PARAMETERS}

Consequently, let us turn to how to connect the soft parameters with
observables.  The essential point is that experimenters measure
kinematical masses, cross sections times branching ratios, electric
dipole moments, etc.  We have to examine how those are expressed in
terms of soft parameters.  We will see that at most two of the soft
parameters can be directly measured, the gluino and gravitino masses.
The gravitino mass can probably only be measured if it is the LSP, and
then only very approximately.  The soft parameter $M_3$ can be deduced
from the gluino mass to about 20\% accuracy from theoretical
uncertainties $\cite{S. Martin}$ associated with large loop corrections
depending on squark masses (not counting experimental uncertainties).

\section{THE CHARGINO SECTOR}

An important case, and the simplest example, is the chargino sector.
The superpartners of $W^{\pm }$ and of the charged Higgs bosons $H^{\pm
}$ are both spin-1/2 fermions and they mix once the electroweak symmetry
is broken, i.e. once the neutral Higgs field get vev's.  There is a
$W\widetilde{W}$ mass term $M_{2}e^{i\phi _{2}},$ a Higgsino mass term
$\mu e^{i\phi _{\mu }}$, and a mixing term, so the chargino mass matrix
is

\begin{eqnarray}
M_{\tilde{C}}=\left( 
\begin{array}{cc}
M_{2}e^{i\phi _{2}} & \sqrt{2}M_{W}\sin \beta \\ 
\sqrt{2}M_{W}\cos \beta & \mu e^{i\phi _{\mu }}
\end{array}
\right) .
\end{eqnarray}

The eigenvalues of this matrix (since it is not symmetric one usually
diagonalizes ${M_{\tilde c}}^\dagger{M_{\tilde c}})$ are the physical mass eigenstates, $M_{
\tilde{C}_{1}}$ and $M_{\tilde{C}_{2}}.$ \ The formulas are a little simpler
if we write them in terms of the trace (sum of eigenvalues) and determinant
(product of eigenvalues),

\begin{eqnarray}
TrM_{\tilde{C}}^{\dagger}M_{\tilde{C}}=M_{\tilde{C}_{1}}^{2}+M_{\tilde{C}_{2}
}^{2}=M_{2}^{2}+\mu ^{2}+2M_{W}^{2},
\end{eqnarray}
\
\
\

$$DetM_{\tilde{C}}^{\dagger }M_{\tilde{C}}=M_{\tilde{C}
_{1}}^{2}M_{\tilde{C}_{2}}^{2}=M_{2}^{2}\mu ^{2}+2M_{W}^{4}\sin
^{2}2\beta$$

\begin{eqnarray}
-2M_{W}^{2}M_{2}\mu \sin 2\beta \cos (\phi _{2}+\phi _{\mu }).
\end{eqnarray}

Several things should be noted $\cite{M. Brhlik}$.  Experimenters will
measure the physical masses $M_{\tilde{C}_{1}}$ and $M_{\tilde{C}_{2}}$.
But what we need to know for the Lagrangian are $M_{2},\mu ,$ and the
phases.  We also need to measure $\tan \beta .$ The phases enter here
only in the combination $\phi _{2}+\phi _{\mu }.$ This is a physical
phase that cannot be rotated away, a true observable, as much as the CKM
phase of the quark mixing matrix.  Such ``reparameterization invariant''
combinations of phases enter all observables.  In general we think of
phases as leading to CP-violating phenomena, and they do, but note that
here the (non-CP-violating) masses also depend strongly on the phases.

After diagonalizing this matrix, the symmetry eigenstates can be
expressed in terms of the mass eigenstates, which will be linear
combinations whose coefficients are the elements of the eigenvectors of
the diagonalizing matrix.  These coefficients also depend on $\tan \beta
$ and the phases, and enter the Feynman rules for producing the mass
eigenstates.  Thus the cross sections and decay branching ratios (BR)
also depend on the phases and $\tan \beta .$

To measure any of the parameters, such as $\tan \beta $, it is necessary
to invert the equations and measure all of them.  Since there are four
parameters here one has to have at least four observables.  In practice
more observables will be necessary since there will be quadratic and
trigonometric ambiguities, and since experimental errors will lead to
overlapping solutions.  Thus we learn the important result
$\cite{M. Brhlik}$ that from the masses alone we cannot measure $\tan
\beta $.

It is worth elaborating on this point, because many papers and
phenomenological analyses have claimed that tan$\beta$ can be measured
in various sectors.  Whenever that is done (except at a lepton collider
with polarized beams or by combining a variety of Higgs sector data ---
see below) the analysis has actually assumed various soft terms are zero
or equal, to reduce the number of parameters.  While such assumptions
may (or may not) be good guesses, once there is data it is necessary to
measure such parameters without assumptions.

The next thing to try is to add the (presumed) cross section data.  The
dominant processes are s-channel Z and $\gamma $, and sneutrino
exchange, as shown (they are shown for $e^+$, $e^-$ beams --- similar
diagrams can be drawn for hadron colliders):

\begin{center}
\includegraphics[scale=.69]{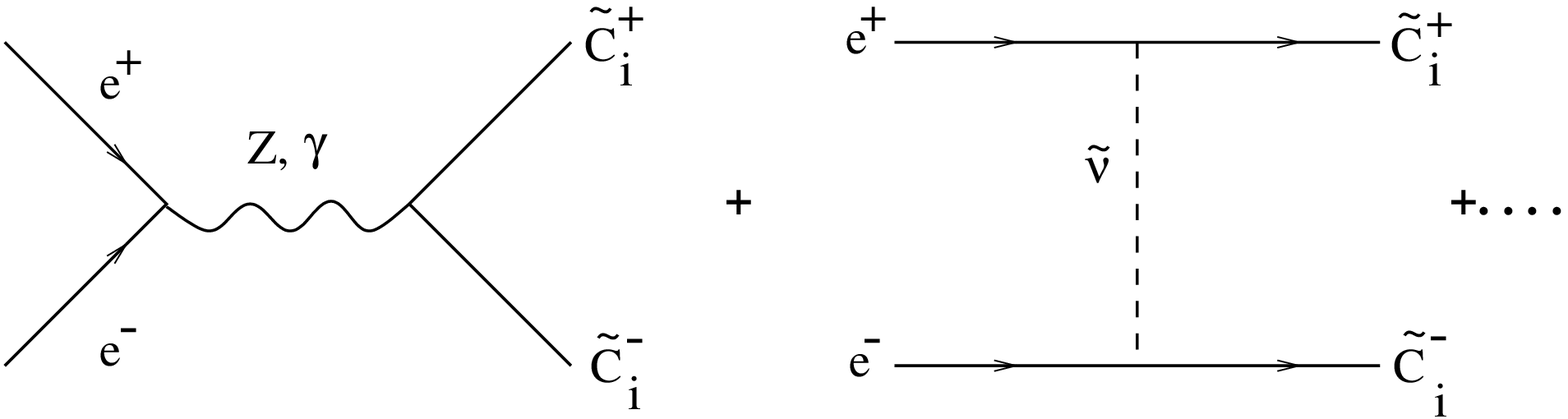}
\epsfxsize=2.7in
\epsfysize=1.25in
\vspace{.5in}
 \hspace*{1in}
\end{center}  


The couplings to Z and $\gamma $ are determined by the diagonalized mass
matrix, but now the sneutrino masses and couplings enter.  If we make no
assumptions there are three sneutrino masses, each with its own
coupling, so 6 new parameters; actually, the sum of squares of the
couplings is determined by supersymmetry to be equal to the $\nu eW$
coupling, so there are only 5 new parameters.  If we don't consider
chargino decays, there are three cross sections,
$\tilde{C}_{1}\tilde{C}_{1},\tilde{C}_{2}\tilde{C}_{2},
\tilde{C}_{1}\tilde{C}_{2}.$ In principle, one can imagine measuring
differential cross sections, obtaining several angular bins.  In
practice, with limited statistics and backgrounds, usually even at an
electron collider one only measures forward-backward asymmetries
$\cal{A}$.  At a hadron collider it would be very hard to measure even
the asymmetries (because of difficulties in reconstructing the
superpartners from their decay products, and because of large
backgrounds), and before they were included in the counting a careful
simulation would have to be done.  Thus, if the produced charginos can
be reconstructed, it may be possible to measure $\tan \beta $ at an
electron collider $\cite{A number of studies}$, but probably still not
at a hadron collider.  However, it needs to be shown that the produced
charginos can be reconstructed even at a lepton collider.

Further, the charginos of course decay. There are a number of possible
channels, some of which are shown:

\begin{center}
\includegraphics[scale=.8]{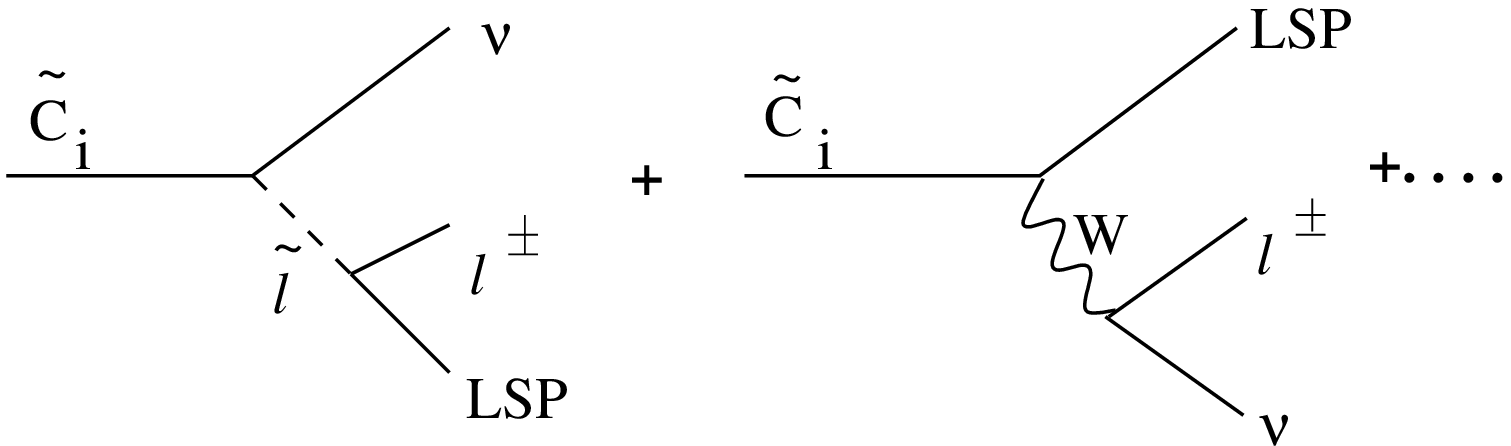}
\vspace{.1in}
\hspace{.3in}
\end{center}


These introduce new parameters, slepton and squark masses and couplings, and
the LSP mass and couplings, even assuming the decay to the LSP dominates
over decay cascades through other neutralinos.  Unless one decay
dominates, too many parameters may enter to measure $\tan \beta $ from
these channels even at a higher energy lepton collider.  

If the decay via an intermediate $W$ dominates, some final polarization can
be obtained, but if sleptons and squarks are light and contribute to the
decays then no polarization information is transmitted to the final state
because they are spinless.  Their chirality can still enter since the
wino component of charginos couples to LH sferm
ions, while the Higgsino
component couples to RH sfer
mions.

In general then it is not possible to measure $\tan \beta $ or the soft
phases or other soft parameters from chargino channels alone, though if
squarks and sleptons are heavy or if charginos can be reconstructed
experimentally it may be possible $\cite{A number of studies}$.  If one
assumes values for phases or assumes relations for parameters the
results for tan$\beta$ and other parameters are not true measurements
and may not correspond to the actual values.  However, if one restricts
the assumptions to highly plausible ones it is worthwhile to make them
and learn as much as possible.  Assumptions that are highly plausible
include assuming that the three sneutrinos are approximately degenerate,
that $\tilde{e}_{L},\tilde{\mu}_{L,}\tilde{\tau} _{L}$ are approximately
degenerate and similarly $\tilde{e}_{R},\tilde{\mu}
_{R},\tilde{\tau}_{R}$ are approximately degenerate, with similar
assumptions for the squarks.  Also, for collider physics the first two
families can be taken to have small L-R mixing, since L-R mixing is
expected to be proportional to the mass of the associated fermions and
thus small.  Under these assumptions it will definitely be possible to
measure $\tan \beta $ and the soft phases at lepton colliders that can
produce some of the superpartners, when the extra observables from beam
polarization and a second energy are included, even if the collider does
not have enough energy to produce many superpartners (see Section 14).
With such assumptions it may even be possible to measure tan$\beta$ and
some phases at hadron colliders.  Some of the assumptions can be checked
independently.  I will call these approximate approaches ``reasonable
approximate models''.

We have only looked at the chargino channels to have a simple example,
but of course all the accessible superpartners will be produced at any
collider, leading to more parameters and more observables.  We will see
below how the counting works for a few kinds of processes, and how
channels can be combined.  Only with good simulations (or of course real
data) can one be confident about counting observables.  My estimates are
that with hadron colliders true measurements of $\tan \beta $ and soft
phases and other soft parameters are not possible, but they may be
possible for reasonable approximate models depending on the actual
values of the parameters.  For lepton colliders with a polarized beam,
above the threshold for some superpartners, the parameters of ${\cal
L}_{soft}$ can be measured, as discussed further below.

\section{NEUTRALINOS}

If charginos are produced, neutralinos will be produced too.  So we get
more observables (masses, cross sections, asymmetries).  There are more
parameters too, but not as many new parameters as new observables.  The
symmetric neutralino mass matrix is

\begin{eqnarray}
M_{\tilde{N}}=\left( 
\begin{array}{cccc}
M_{1}e^{i\phi _{1}} & 0 & -M_{Z}\sin \theta _{W}\cos \beta & M_{Z}\sin
\theta _{W}\sin \beta \\ 
& M_{2}e^{i\phi _{2}} & M_{Z}\cos \theta _{W}\cos \beta & -M_{Z}\cos \theta
_{W}\sin \beta \\ 
&  & 0 & -\mu e^{i\phi _{\mu }} \\ 
&  &  & 0
\end{array}
\right) ,
\end{eqnarray}

in the symmetry basis ($\tilde{B},\tilde{W},\tilde{H}_{U},\tilde{H}_{D})$. 
Even when the elements are complex it can be diagonalized by a single
unitary matrix.  

We saw that the chargino sector depended on a single physical phase,
reparameterization invariant, $\phi _{2}+\phi _{\mu }.$  Similarly, there
are two physical phases, that cannot be rotated away, in the neutralino
matrix.  One can see this various ways --- one is by simply calculating
observables.  A somewhat easier one is by redefining the basis by

\begin{eqnarray}
\left( 
\begin{array}{cccc}
e^{-i\phi _{\mu }/2} &  &  &  \\ 
& e^{-i\phi _{\mu }/2} &  &  \\ 
&  & e^{-i\phi _{2}/2} &  \\ 
&  &  & e^{-i\phi _{2}/2}
\end{array}
\right) ,
\end{eqnarray}

in which case the resulting matrix depends explicitly only the
physical phases.

So there is one new soft mass, $M_{1}$, and one new physical phase, $\phi
_{1}+\phi _{\mu }.$ In principle the masses of the four mass eigenstates
can be measured, and the cross sections
$\tilde{N}_{1}+\tilde{N}_{1},\tilde{N}_{1}+
\tilde{N}_{2},\tilde{N}_{2}+\tilde{N}_{2},$etc. and associated
asymmetries.  The number of new observables is different at different
colliders.  Below we will also discuss the number of parameters and
observables from the Higgs sector explicitly.  It is extremely important
for detector groups at various colliders to count the number of
observables they can expect to measure.  This has to be done using
models, of course, so it is very important that the models be ones that
can describe the electroweak symmetry breaking without excessive fine
tuning.  Of course, the models should also be consistent with LEP data.

\section{SOME EFFECTS OF PHASES FOR THE HIGGS SECTOR -- TRUE LEP LIMITS}

Next let us turn to examining how the phases affect the physics of the Higgs
sector $\cite{M. Brhlik,A. Pilaftsis,K. S. Babu,G. L. Kane}$.  It is important to realize that the phases that enter the Higgs
sector are not new ones, they are some of the same ones that are already in $
{\cal L}_{soft}$.  At tree level it has long been understood that all the
quantities that affect the Higgs physics can be chosen to be real.  The
phase effects enter at one loop order, because the stop loops are a
large contribution$\cite{M. Brhlik,A. Pilaftsis}$. The stop
mass matrix is in general complex, just as the chargino mass matrix is complex, and so
the phases enter into the scalar effective potential.  

One can write

\begin{eqnarray}
\hspace{.45in}H_{D}=\frac{1}{\sqrt{2}}\left( 
\begin{array}{c}
\text{v}_{D}+h_{1}+ia_{1} \\ 
h_{1}^{-}
\end{array}
\right),\nonumber\\
H_{U}=\frac{e^{i\vartheta }}{\sqrt{2}}\left( 
\begin{array}{c}
\text{v}_{U}+h_{2}+ia_{2} \\ 
h_{2}^{+} \\ 
\end{array}
\right) ,
\end{eqnarray}

with the vev's taken to be real, and $\tan \beta =$v$_{U}/$v$_{D}$.  The
phase $\vartheta $ allows a relative phase between the two vev's at the
minimum of the Higgs potential.  The stop mass matrix has off diagonal L-R
elements

\begin{eqnarray}
\hspace{.2in}y_{t}(A_{t}H_{U}-\mu ^{\ast }H_{D}^{\ast }),
\end{eqnarray}

so the phases of the triscalar coupling $A_{t}$ and of $\mu $ and the
relative phase $\vartheta $ enter the stop mass eigenvalues.  The scalar
potential contains a term

\begin{eqnarray}
\hspace{.1in}\sum m_{\tilde{t}_{i}}^{4}\ln m_{\tilde{t}_{i}}^{2},
\end{eqnarray}

and the phases $\phi_{A_{t}},\phi _{\mu}$ enter through the stop mass
eigenvalues $m_{\tilde{t}_{i}}.$ The scalar potential $V$ is then
minimized by setting $\partial V/ \partial h_{1},\partial V/ \partial
h_{2},\partial V/$ $\partial a_{1}, \partial V/ \partial a_{2}=0$.
Two of the resulting four equations are not independent, so three
conditions remain.

The Higgs sector then has 12 parameters, v$_{U},$v$_{D},$ $\phi _{A_{t}},\phi
_{\mu },\vartheta ,A_{t},\mu ,$ the parameters $m_{\tilde{Q}
}^{2},m_{\tilde{U}}^{2},b,m_{H_{U}}^{2},m_{H_{D}}^{2}$ from ${\cal L}_{soft}$, and
the renormalization scale $Q$ since the parameters run.  Three can be
eliminated by the three equations from minimizing $V$.  The scale $Q$ is
chosen to minimize higher order corrections.  The conditions that guarantee
EWSB occurs let us replace v$_{U},$v$_{D}$ by $M_{Z}$ and $\tan \beta $ as
usual.  Thus there are 7 physical parameters left, including $\tan \beta $
and one physical (reparameterization-invari
ant) phase $\phi _{A_{t}}+\phi _{\mu }.$  This number cannot be
reduced without new theoretical or experimental information. Any
description of the Higgs sector based on fewer than 7 parameters has made
arbitrary guesses for some of these parameters, and is likely to be wrong. 
If $\tan \beta $ is large, then sbottom loops can also enter $V$ and
additional parameters are present.

If the phase is non-zero it is not possible to separate the ``pseudoscalar'' 
$A=\sin \beta a_{1}+\cos \beta a_{2}$ from $h,H$ so it is necessary to
diagonalize a 3x3 mass matrix.  We name the three mass eigenstates $H^{i}$;
in the limit of no CP-violating phase $H^{1}\rightarrow h, H^{2}\rightarrow
A,H^{3}\rightarrow H.$  Then, of course, all three mass eigenstates can
decay into any given final state or be produced in any channel, so there
could be three mass peaks present in a channel such as $Z+$Higgs (wouldn't
that be nice).  All production rates and branching ratios depend on the
phase, and can change significantly.

\vspace{.2in}
\begin{center}
\includegraphics[width=4.2in]{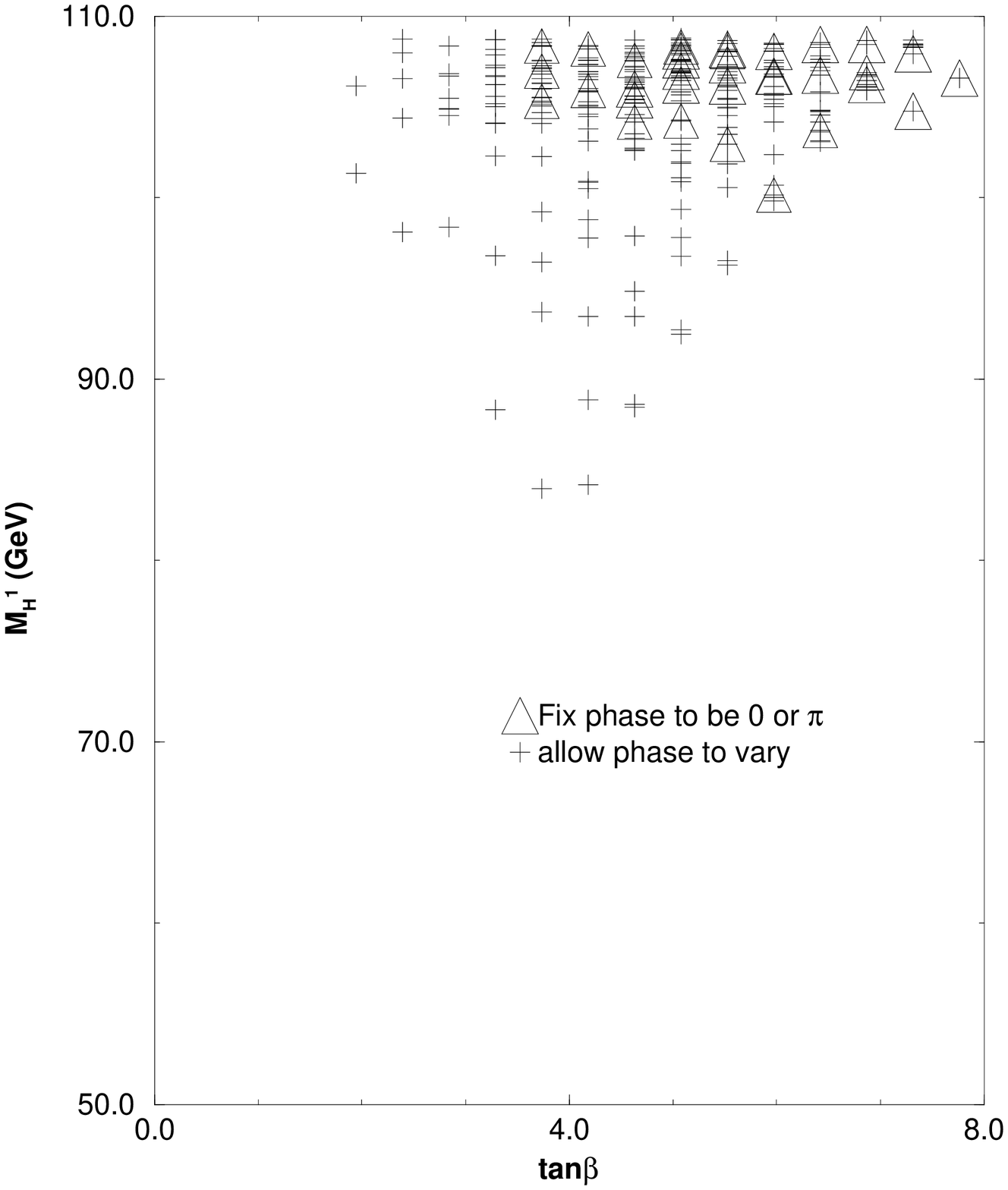}
\end{center}
\hspace{2.5in} Figure 7.1

It is particularly interesting to ask two questions $\cite{G. L. Kane}$.  (1) Suppose there is no Higgs
boson found at LEP.  Then there is an experimental limit on $\sigma
(H^{1}) \times BR(H^{1} \rightarrow b\bar{b}).$  What are the lower limits on $
m_{H^{1}}$and on $\tan \beta $ in the full 7 parameter theory compared to
the SSM with the phase set to $0$ or $\pi $?  This is answered in Figure
7.1.  We see that the mass of the lightest Higgs boson is allowed to be
significantly lighter if the phase is present.  To summarize, the lower limit for the SSM without phases is about 10\% below
the SM limit, and the lower limit with the phase present is an additional
10\% lower (about 85 GeV).  Similarly, $\tan \beta $ can take on lower
values if the phase is non-zero, probably values down to about 1.5 or
even below, so very little of the perturbatively allowed range of
$\tan \beta $ is excluded, contrary to what is often asserted by LEP
experimental groups or theorists analyzing the data.


\vspace{.2in}
\begin{center}
\includegraphics[width=4.2in]{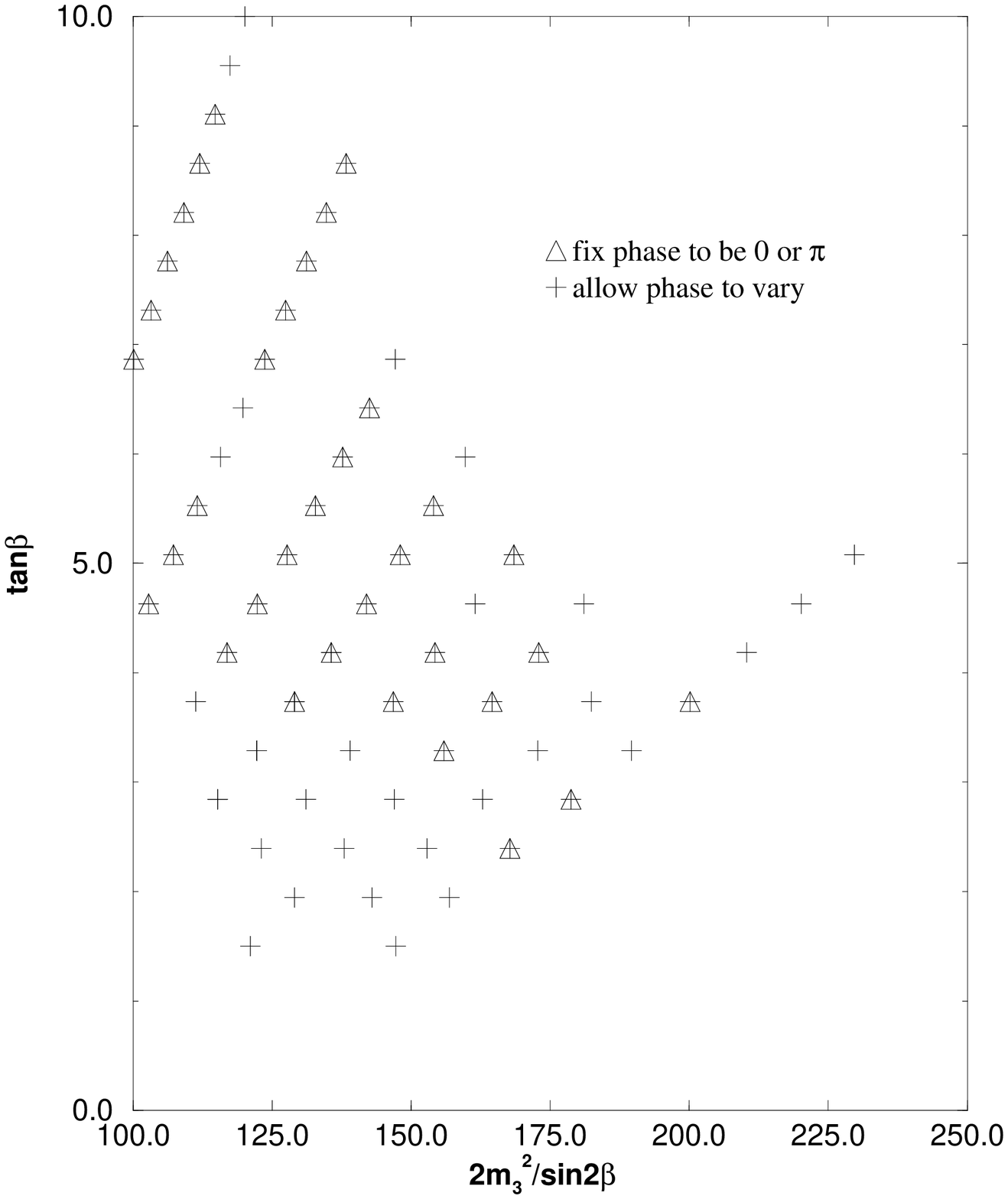}
\end{center}
\hspace{2.5in} Figure 7.2

(2) Suppose a Higgs boson is found.  Then $m_{H^{1}}$and its $\sigma
\times BR$
are measured.  What region of the full 7 parameter space is allowed? 
Again the answer is different for the case with phase set to $0,\pi $ and
the case with general phase allowed, as shown in Figure 7.2.  Thus once there
is a discovery it will be seriously misleading to not include the phase in
the analysis.  The quantity $2m_{3}^{2}\sin 2\beta $ plotted on the horizontal
axis is what would be $m_{A}^{2}$ if the phase were $0,\pi .$ 

In both cases the results shown here are simply meant to illustrate the
effect of the phase.  The full range of the other parameters is not
covered, and experimental aspects are not included except for crude
estimates of efficiencies.  The proper analyses need to be done by
experimenters.  If the heavier Higgs bosons are heavy, so that they
decouple, the effect decreases for the lower limit on the mass of the
lightest eigenstate, but there is still a significant effect on the lower
limit on $\tan \beta $.


With 7 parameters we would need at least 7 observables if we were to determine (say) $
\tan \beta $ from the Higgs sector alone. The possible observables are the
3 neutral scalar mass eigenstates, the charged Higgs mass, the three $\sigma
 \times BR$ for $Z+$Higgs and three $\sigma \times BR$ for channels $H^{i}+H^{j}$, and
the two stop mass eigenstate masses.  Probably in addition one can add the
ratio $r=\sigma (gg\rightarrow H^{2}\rightarrow b\bar{b})/\sigma
(gg\rightarrow H^{1}\rightarrow b\bar{b})$. \ So eventually it will be
possible to measure $\tan \beta $ and the phase $\phi _{A_{t}}+\phi _{\mu }$
in the Higgs sector and compare with the $\tan \beta $ measured in the
gaugino sector.  LHC data may allow one to carry out such an analysis.

\section{THE GLUINO PHASE}

The phase of the gluino or SU(3) gaugino nicely illustrates some of the
subtleties of including and measuring the phases $\cite{S. Mrenna}$.  People have wondered if
it could be rotated away, or if it was not measurable.  More precisely,
only certain combinations of phases are invariant under phase
redefinitions, as we saw above and as we will see further
below.  The gluino phase $\phi_3$ is not itself a
physical observable phase.  In these lectures we will not discuss this
issue in detail, but simply find observable sets of phases in pratical
situations.  The general analysis is understood, and can be found in
refrences $\cite{S. Th,G. Good}$.  We will see how to
measure $\phi_3$ (relative to other phases) below.  In the two component notation that is often used in the
supersymmetry formalism, the gluino part of the Lagrangian is

\begin{eqnarray}
\hspace{.25in}{\cal L}=-\frac{1}{2}(M_{3}e^{i\phi _{3}}\lambda \lambda +M_{3}e^{-i\phi _{3}}
\bar{\lambda}\bar{\lambda})
\end{eqnarray}

where $\lambda $ is the gluino field.  It is inconvenient to have complex masses in the
Feynman rules, so we redefine the field,

\begin{eqnarray}
\hspace{.15in}\psi=G\lambda ,\bar{\psi}=G^{\ast }\bar{\lambda}
\end{eqnarray}

where $G=e^{i\phi _{3}/2}$.  Then for any flavor the Feynman rules
introduce factors of $G$ or $G^{\ast }$ at the vertices,

\begin{center}              
\includegraphics[scale=.69]{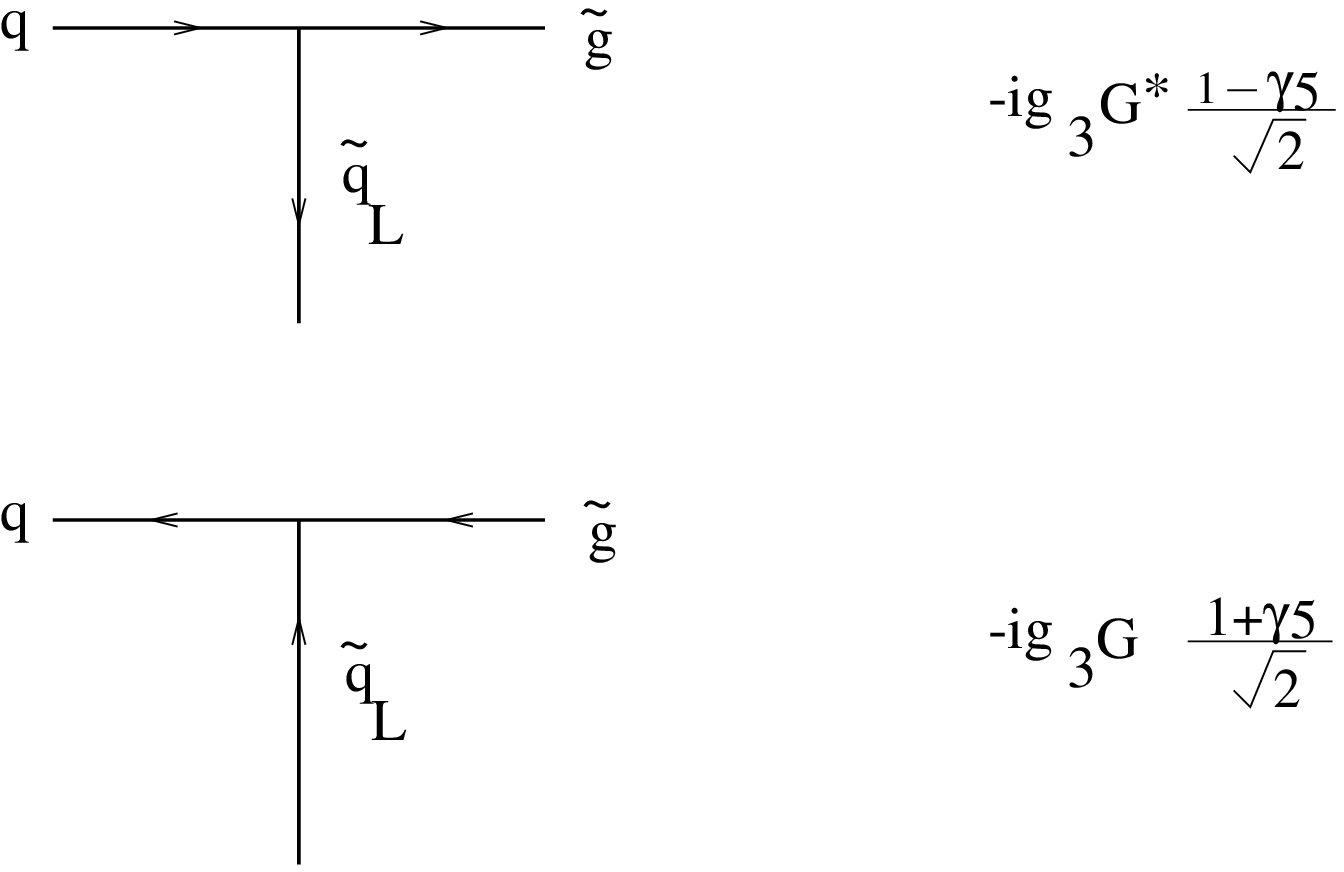}
\end{center}


not including the color factors.  

Now consider gluino production $q+\bar{q}\rightarrow \tilde{g}+\tilde{g}$. 
Factors of $G$ and $G^{\ast }$ enter so that there is no dependence on
the phase from these two diagrams.

\begin{center}
\includegraphics[scale=.8]{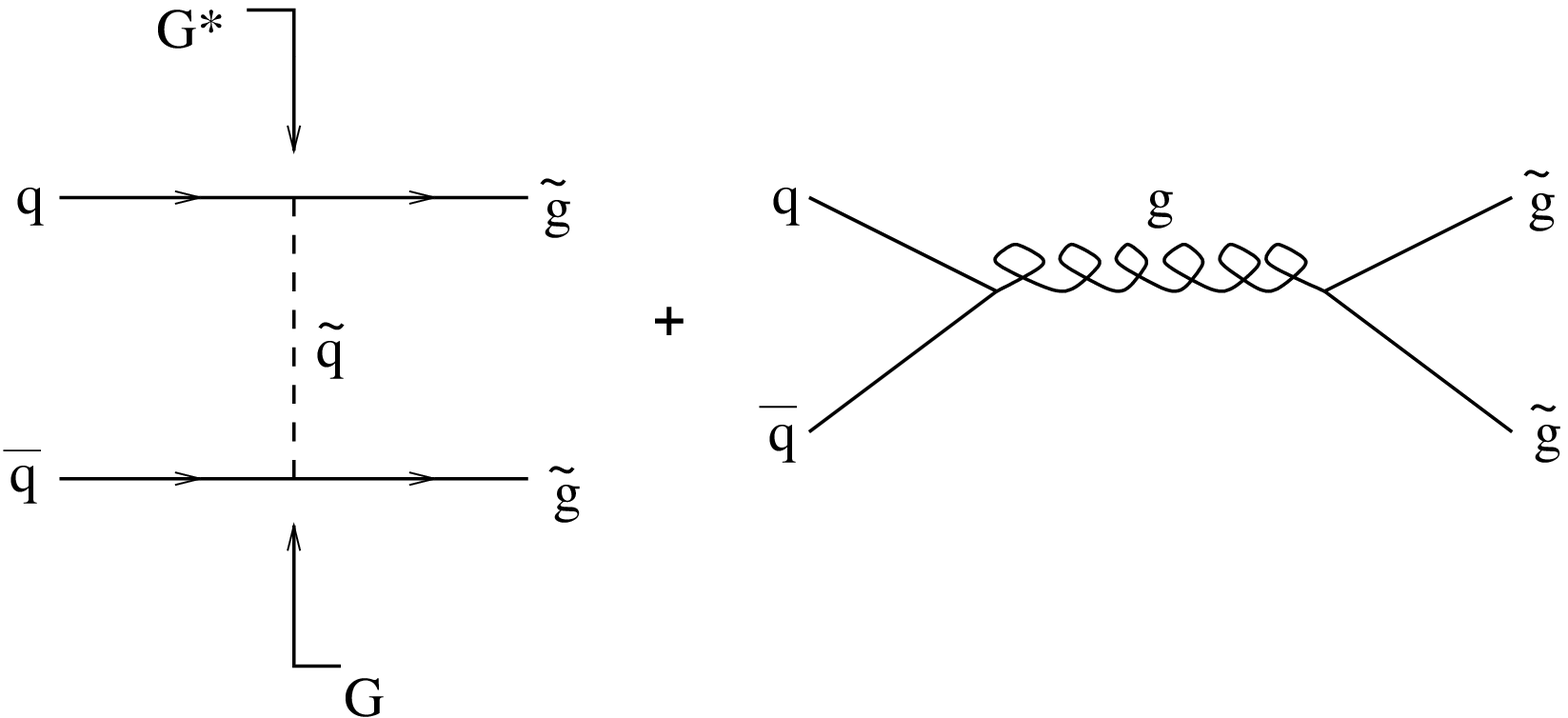}
\end{center}

If we consider $q+g\rightarrow \tilde{q}+\tilde{g},$ production of $\tilde{q}
_{L}$ leads to an overall factor of $G^{\ast },$ while production of $\tilde{
q}_{R}$ gives an overall $G$.  The phase is then observable if $L$-$R$
mixing occurs, but $L$-$R$ mixing is expected to be very small for the first
two families (because it is proportional to the off-diagonal element of
the sfermion mass matrix, which is usually taken to be proportional to
the fermion mass), and those constitute the beams we have to use.  Thus
gluino production above does not seem to depend on the phase. 

But gluinos have to decay, and then the phases enter.  For simplicity,
imagine the gluino decay is via a squark to $q\bar{q}\tilde{\gamma},$ as
shown for $\tilde{q}_{L},$

\begin{center}
\includegraphics[scale=.8]{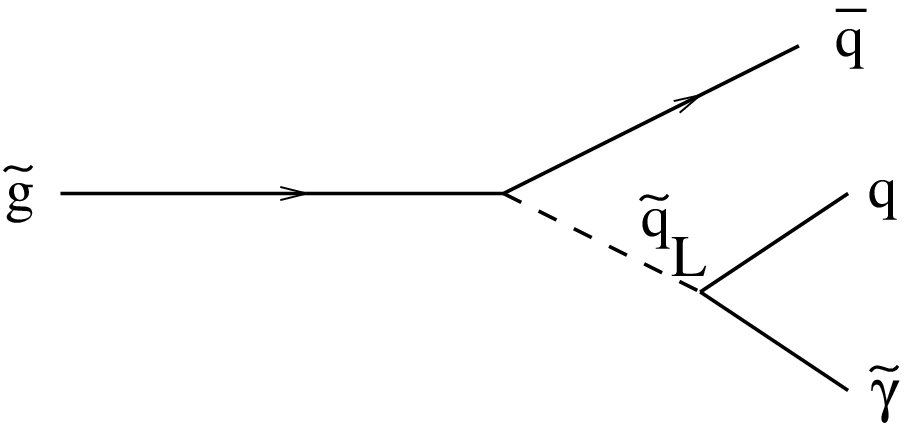}
\end{center}


Then a factor $e^{i\phi _{3}/2}$ enters at the gluino vertex, and a factor $
e^{-i\phi _{1}/2}$ at the photino vertex.  The resulting differential cross
section is

$$\frac{d\sigma}{dx}=32g_{3}^{2}(ee_{q})^{2}(\frac{1}{m_{\tilde{q}_{L}}^{4}}+
\frac{1}{m_{\tilde{q}_{R}}^{4}}){m_{\tilde{g}}^{4}x\sqrt{x^{2}-y^{2}}}$$
\begin{eqnarray}
(x-\frac{4}{3}x^{2}-\frac{2}{3} y^{2}+xy^{2}+y\left(1-2x+y^{2}\right)
\cos(\phi_{3}-\phi_{1}))
\end{eqnarray}

where
 $x=E_{\tilde{\gamma}}/m_{\tilde{g}},y=m_{\tilde{\gamma}}/m_{\tilde{g}}.$
 Note that what enters is the physical, reparameterization invariant
phase $\phi _{3}-\phi _{1}.$  Reference \cite{S. Mrenna} has studied
how various distributions depend on this phase and on $\tan \beta $ and
the soft masses so measurements can be made at the Tevatron and LHC.  

\section{CP VIOLATION}

We have looked at some examples of how the phases affect masses, cross
sections, and branching ratios (their effects in the Higgs sector involve CP
violating aspects, but somewhat indirectly).  Now let us turn to the unique
effects they have on CP violation.  We will see that soft supersymmetry
phases could explain all the CP violation in the Kaon and B systems.  

How do we expect CP violation to arise in a fundamental way?  In string
theory, Strominger and Witten $\cite{A. Strominger}$ showed in 1985 that in string perturbation
theory CP was a good symmetry, that could be spontaneously broken by moduli
vev's. Then in 1992 Dine, Leigh, and MacIntire pointed out $\cite{M. Dine}$ that 4D CP
transformations are discrete gauge transformations inherited from l0D string theory, so even
non-perturbatively there will be no explicit breaking of CP.  Thus CP
violation can arise from spontaneous breaking that leads to 
(a) complex Yukawa couplings that in turn lead to a CKM
phase, $\delta _{CKM}$, and (b) the phases of ${\cal L}_{soft}.$  

Since the CKM phase cannot explain the baryon asymmetry there must be some
other phases that are important, and the soft phases are a leading 
candidate.  Therefore, it is very interesting to examine the possibility that
the CKM phase is actually very small, and all CP violation arises from the
soft phases.  Since the CKM phase arises from the
supersymmetry-conserving superpotential if the Yukawa couplings have a
relative phase, while the soft phases arise from the supersymmetry-breaking,
it is easy to imagine such a world, and we will examine it
phenomenologically.  The possibility that all CP violation arose from soft
supersymmetry phases $\cite{J.-M. Frere}$ was first suggested by Frere and Gavela in 1983, and
reexamined by Frere and Abel in 1997.

There are currently six pieces of experimental information on CP violation.
 We can compare the SM and a possible version of supersymmetry with
 large flavor-independent phases and significant off-diagonal triscalar couplings:

\begin{tabular}{ccc}\hline
 && \textit{A SUPERSYMMETRY}\\
\textit{DATA} & \textit{SM}& \textit{MODEL} (\textit{and}\;$\delta _{CKM}=0)$ \\\hline
$\varepsilon _{K}$ & OK & OK \\ 
($\varepsilon ^{\prime }/\varepsilon )_{K}$ &  PROBABLY OK & OK, requires\\
&&non-trivial flavor physics \\ 
$\sin 2\beta $ & OK & OK \\ 
upper limits on nEDM, eEDM & very good & OK \\ 
baryon asymmetry & NO & OK \\ 
strong CP problem & NO & ?
\end{tabular}

In the K and B systems all supersymmetry effects are from loops.  For the K
system $\cite{Some of a number,Michal Brhlik}$ the dominant contributions are probably gluino-squark boxes and
penguins.  For $\Delta m_{K}$ the usual SM box is present, and the
supersymmetry boxes could be of the same order or somewhat smaller --- the
hadronic uncertainties are large enough so one cannot tell.  For $
\varepsilon _{K}$ the imaginary part of the box is needed, so with our
working assumption that $\delta _{CKM}$ $\approx 0$ there is no SM
contribution.  The gluino-squark box is

\begin{center}
\includegraphics[scale=.8]{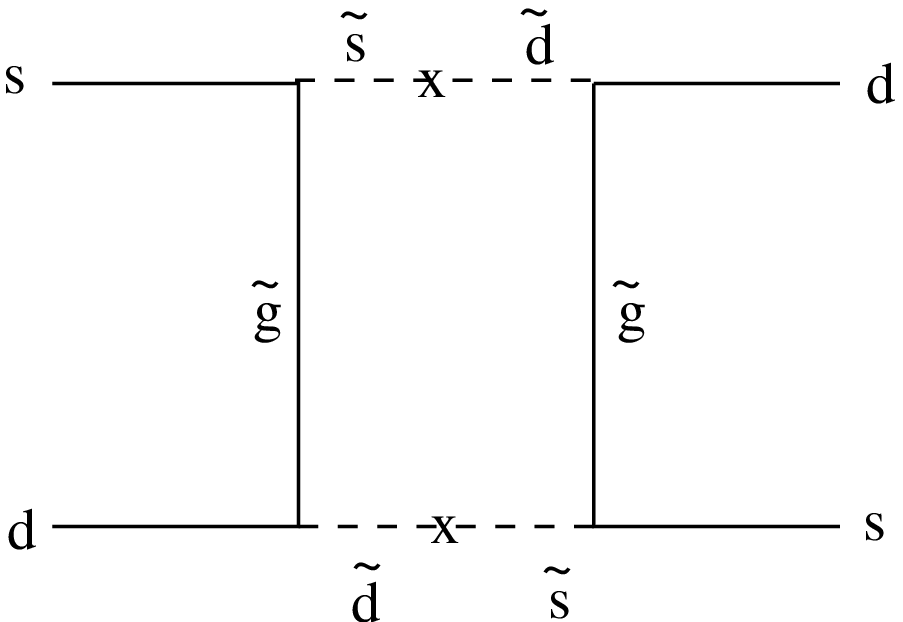}
\end{center}


The ``x'' on a line means a squark L-R chirality flip, so it introduces a
factor proportional to the off-diagonal elements of the trilinear couplings $
A$.  As we saw above, the gluino phase $\phi _{3}$ enters at the
squark-gluino vertex --- this is the same phase that can be studied at
colliders.  The contribution from this diagram is then proportional to

\begin{eqnarray}
\hspace{.25in}Arg(A_{sd}e^{-i\phi _{3}})\left| A_{sd}\right| .
\end{eqnarray}

It is important $\cite{Michal Brhlik}$ that the physical, reparameter
ization-invariant phase
combination $\phi_{sd}-\phi_3$ enters; the gluino phase alone would not
give a meaningful result.  The magnitude of the off-diagonal $sd$ element
of the triscalar couplings must be of the right size to describe $
\varepsilon _{K},$ so the supersymmetry approach can describe it but not
explain it until there is a compelling flavor model.  The SM gives a more
natural description here, with the size of $\varepsilon_K$ given by the
Jarlskog invariant, but recall we are pursuing CP violation beyond the
SM because the SM fails when we get to the baryon asymmetry.

For $\varepsilon ^{\prime }/\varepsilon $ the same sparticles enter in the
penguin diagram,

\begin{center}
\includegraphics[scale=.8]{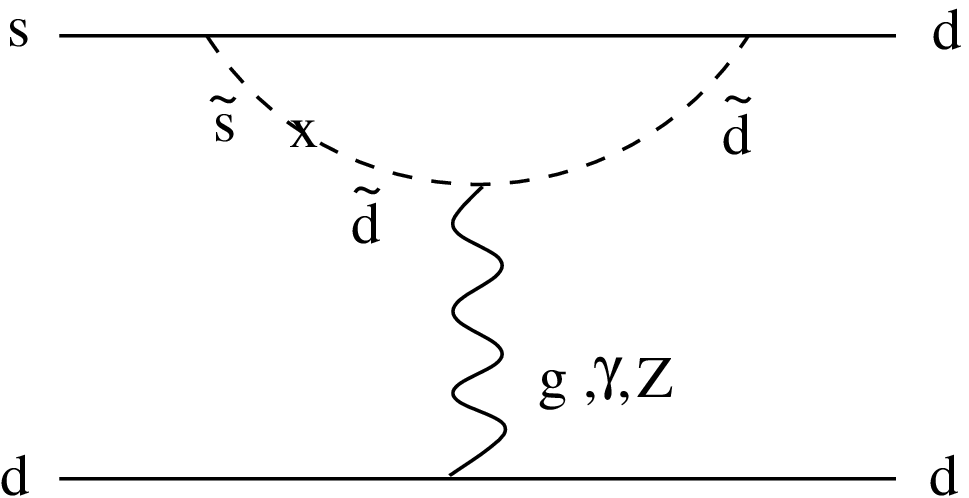}
\end{center}


which is again proportional to the same factor of Eq. 9.1, so
the supersymmetry description is significantly constrained in order to
describe both $\varepsilon $ and $\varepsilon ^{\prime }/\varepsilon .$ 
Results are usually presented in terms of quantities $\delta _{AB}^{f,ij}$,
where $f$ and $i,j$ are flavor indices, and $A,B$ are chirality indices. 
We obtain a consistent description if

\begin{eqnarray}
\left| \delta _{LR}^{d,12}\right|\approx 3\times10^{-3},\;\;\;Arg
(\delta _{LR}^{d,12}M_{3}^{\ast})\approx 10^{-2},
\end{eqnarray}

where the second quantity is proportional to the factor of Eq. 9.1.  

An interesting thing to note here is that the size of the phase in Eq.
9.1 is perhaps reasonable because it can arise naturally in theories,
where relations often occur among soft terms; such a result can indeed
occur in the D-brane model
described below.  In that model $\phi _{\delta _{LR}^{d,12}}=\phi
_{A_{sd}}=\phi _{M_{3}}$, so at the high scale $Arg(\delta M_{3}^{\ast })$
vanishes, and a small value is generated because $\phi _{A}$ and $\phi
_{M_{3}}$ run differently.

Next turn to the B system $\cite{Michal Brhlik}$.  Here all decays (except $b\rightarrow
s\gamma )$ have a tree level contribution, so (with the assumption that
$\delta _{CKM}\approx 0)$ we expect the B system to be superweak, i.e.
all effects on the B system arise from mixing effects, $\varepsilon
_{B}^{\prime }\approx 0.$ The dominant mixing is generated by the
chargino-stop box,

\begin{center}
\includegraphics[scale=.8]{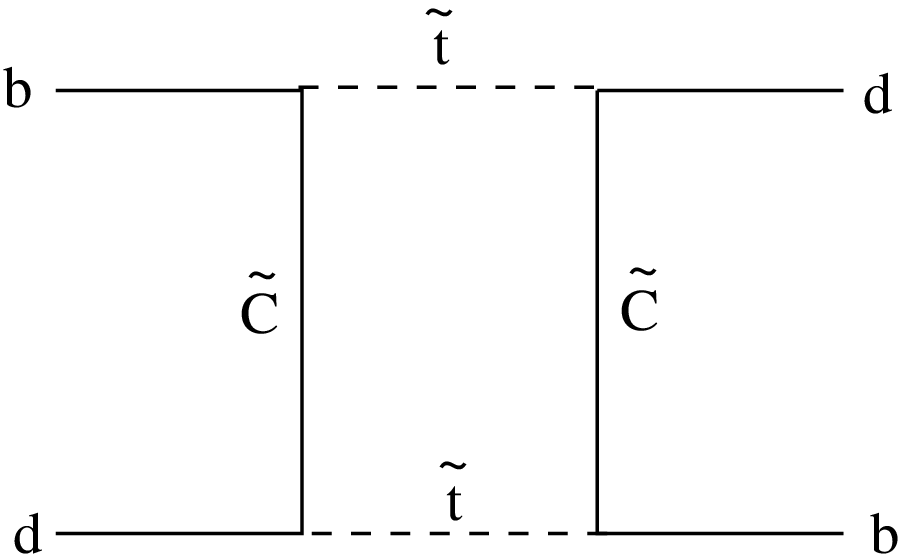}
\end{center}


and the phases $\phi _{2},\phi _{\mu },\phi _{A_{t}}$ enter at the vertices.  With our
assumptions we predict (sin 2$\alpha$ and sin 2$\beta$ are defined to be the CP
asymmetries measured in the decays $B_d \rightarrow \pi^{+}\pi^{-}$ and
$\psi K_s$)

\begin{eqnarray}
\hspace{.15in}\sin 2\alpha =-\sin2\beta ,
\end{eqnarray}

since the decay phase is zero.  We also expect $\left| \sin \beta /\sin
\alpha \right| =1$ (this does not uniquely follow from the previous
equation, but in our model it does), while in the SM $\left| \sin \beta
/\sin \alpha \right| =\left| V_{ub}/V_{cb}\sin \vartheta _{C}\right| \leq
0.45.$  Note that these results do not depend on any of the soft
parameters, so they are general tests of the whole approach.  Numerical
studies show that there are regions of parameter space where this model gives a
value of $\sin 2\beta $ consistent with $\Delta m_{B_{d}}$ and recent
measurements.  If $\delta_{CKM}$ is not small we have to do a more
detailed quantitative study, which is underway.

The CP asymmetry for $b\rightarrow s\gamma $ is particularly interesting
because there is no tree level contribution.  The SM CP asymmetry is very
small $\cite{A. Kagen}$, about half a percent.  The supersymmetric loops can
be of the same order as the SM loops (subject to the constraint that the BR
is consistent with data), so a much larger CP asymmetry is possible. 
Simple estimates give several percent, and 10-15\% is possible in
special cases.  This measurement is relatively clean since one is looking
at a charge asymmetry, and many systematic problems cancel in ratios.  It
could be the first place in meson decays where physics beyond the SM is
found.  A 2\% measurement would require perhaps 3000 events of each sign,
which is reasonable for the B factories.

\section{BARYOGENESIS}

One of our most important clues to physics beyond the SM is the fact that
the universe is made of baryons and not antibaryons.  It is known that CP
violation is a necessary condition to explain that asymmetry, and also that
the SM cannot explain the baryon asymmetry $\cite{M. B. Gavela}$, whatever the value of $\delta
_{CKM}$.  Several approaches are known $\cite{Some recent articles}$ that seem to be capable of
explaining the baryon asymmetry, though in all cases the analysis is very
complicated and the resulting numbers are still very uncertain.  One
attractive way to generate the baryon asymmetry as the universe cools is at
the EW phase transition.  Although such non-perturbative effects are very
difficult to calculate, it is thought that the effects of EW sphalerons are
strong enough to wash out any baryon asymmetry generated at higher
temperatures, until the universe has cooled below the temperature at which
the EW symmetry is broken.  That leads many people to expect that the
asymmetry generated at the EW phase transition is the relevant one.

Several other ways to generate the asymmetry could be relevant. One
is the Affleck-Dine mechanism. Energy stored in scalar fields at high
temperatures is radiated into particles whose decay into baryons is
asymmetric.  The second is leptogenesis from decay of heavy Majorana
neutrinos, typically of mass of order 10$^{11}$ GeV. In theories that
conserve B$-$L, sphaleron interactions will generate a baryon asymmetry
from the lepton number violating Majorana neutrino decay. 
In both cases it is very interesting to examine the origin of the needed CP
violating phase.  In the first it is likely to be connected to the
supersymmetry soft phases since the scalars typically have flat potentials
until the supersymmetry is broken.  In the leptogenesis case the origin of the phase is seldom
discussed.  One can imagine that it occurs in the couplings of the Majorana
neutrinos in the superpotential, Yukawa couplings or higher dimension
operators, or in soft terms involving the Majorana fields. Further work is
needed to understand this important question.  Another possibility is
GUT baryogenesis preserved by B-L conservation and involving GUT Yukawa phases.

When the baryon asymmetry is generated at the EW phase transition, it
involves the interactions of charginos, neutralinos, and stops inside and
outside of growing bubbles where the EW symmetry is broken.  If these
sparticles are light enough, which essentially means lighter than $m_{top}$,
an asymmetry will be generated.  It will be proportional to quantities such
as $\sin (\phi _{2}+\phi _{\mu }),\,\,\sin (\phi _{1}+\phi _{\mu }),$ and
other such phases including $\phi _{A_{t}}.$ There are a number of
uncertainties in the calculations, and perhaps the most difficult is how
much of the asymmetry is left after all the sphaleron interactions become
unimportant.  In any case, learning the origin of the CP-violating phase
responsible for baryogenesis is a central issue in particle physics,
with many possible implications from string theory to collider
phenomenology and to cosmology.

\section{IS THE LSP THE COLD DARK MATTER?}

Let us assume that one day superpartners are found at colliders, and the LSP
escapes the detectors.  In addition, WIMP signals are seen in the
underground detectors (DAMA, CDMS, and others), and perhaps in other large
underground detectors and space-based detectors.  Has the cold dark matter
(CDM) of the universe been observed?  Maybe, but those signals don't
demonstrate that.  The only way $\cite{D. J. Chung}$ to know if the CDM has been detected is to
calculate its contribution $\Omega _{LSP}$ to the relic density $\Omega $,
and show that $\Omega _{LSP}\approx 0.3.$  In fact, to some extent a
large scattering cross section, which make direct detection easier, is
correlated with a large annihilation cross section, which reduces the relic
density. 

\vskip -25mm
\begin{center}
\includegraphics[scale=.6]{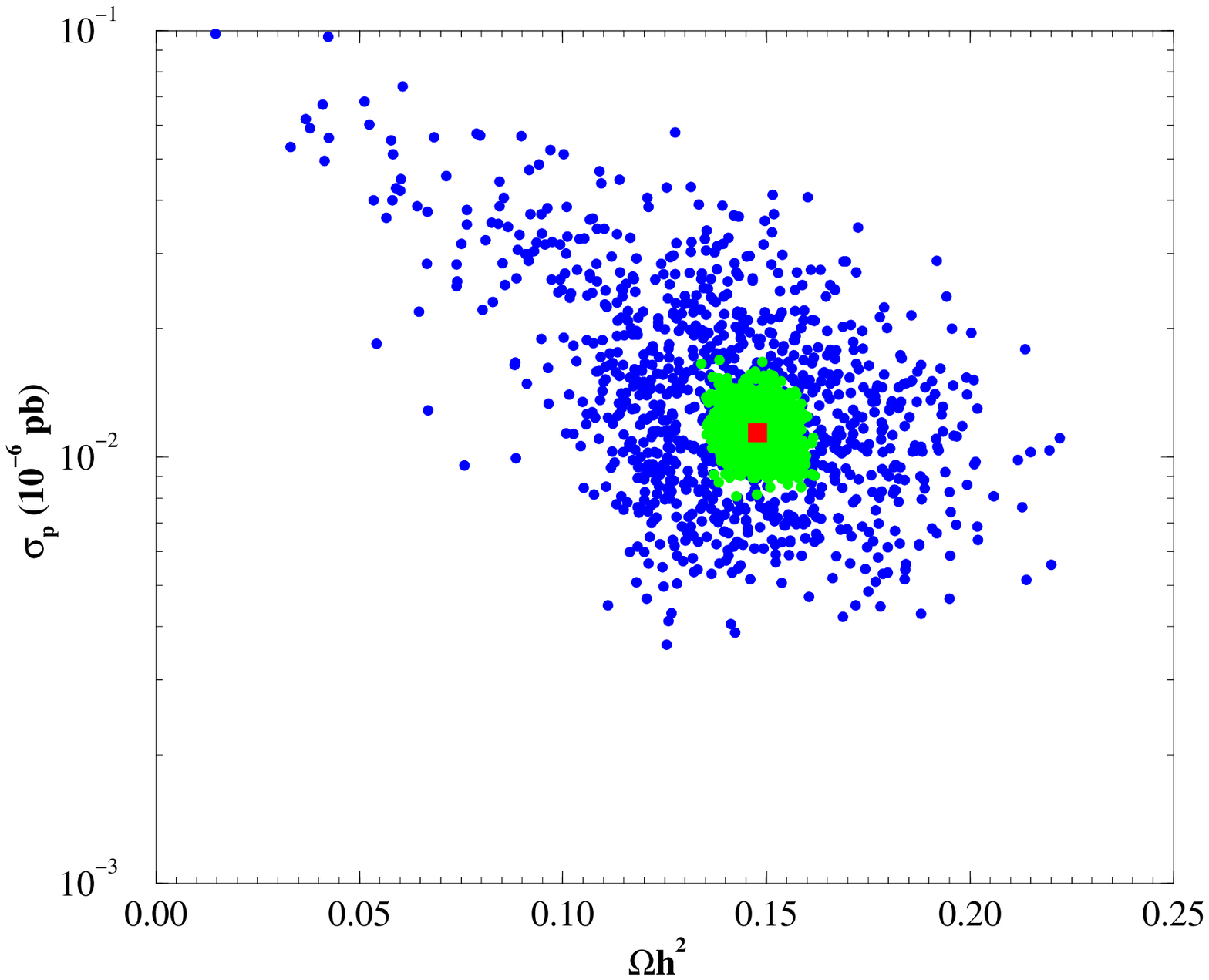}
\end{center}
\hspace{2.5in}Figure 11.1


This subject is relevant in
these lectures because the calculation of the relic density depends on
knowing $\tan \beta $ and various soft parameters, including some of the phases. 
We expect the LSP to be the lightest neutralino (though the gravitino is
a possibility, as are sneutrinos if the LSP relic density is small; in
these lectures we assume conservation of R-parity or an equivalent
quantum number).  Recalling the discussion
of neutralinos above, the lightest eigenvalue of the neutralino mass matrix
will be a linear combination of the four neutralino symmetry eigenstates. 
The coefficients in the ``wave function'' that specifies the linear
combination each depend on the entries in the neutralino mass matrix, $M_{1},M_{2},\mu ,\phi _{1}+\phi _{\mu },\phi _{2}+\phi _{\mu },$and $\tan
\beta $.  Unless all of these are measured it is not possible to
reliably calculate
the relic density. The eigenvectors of the mass matrix determine the
couplings of the LSP to particles that will appear in the annihilation
diagrams.  Additional soft parameters can enter.  For example, the
annihilation of LSP's that largely determines how many remain as the
universe cools can proceed mainly through stau exchange, in which case the
stau mass and its coupling to the LSP have to be measured. The phases can
be very important, for several reasons. In addition to making it harder to measure 
$\tan \beta $, they can also cause explicit variation of $\Omega _{LSP}$ as we
will see in the figures$,$ and they can shift the normal P-wave suppressed
annihilation to helicity-suppressed S-wave annihilation since they mix final
states with different CP.

\begin{center}
\includegraphics[scale=.55]{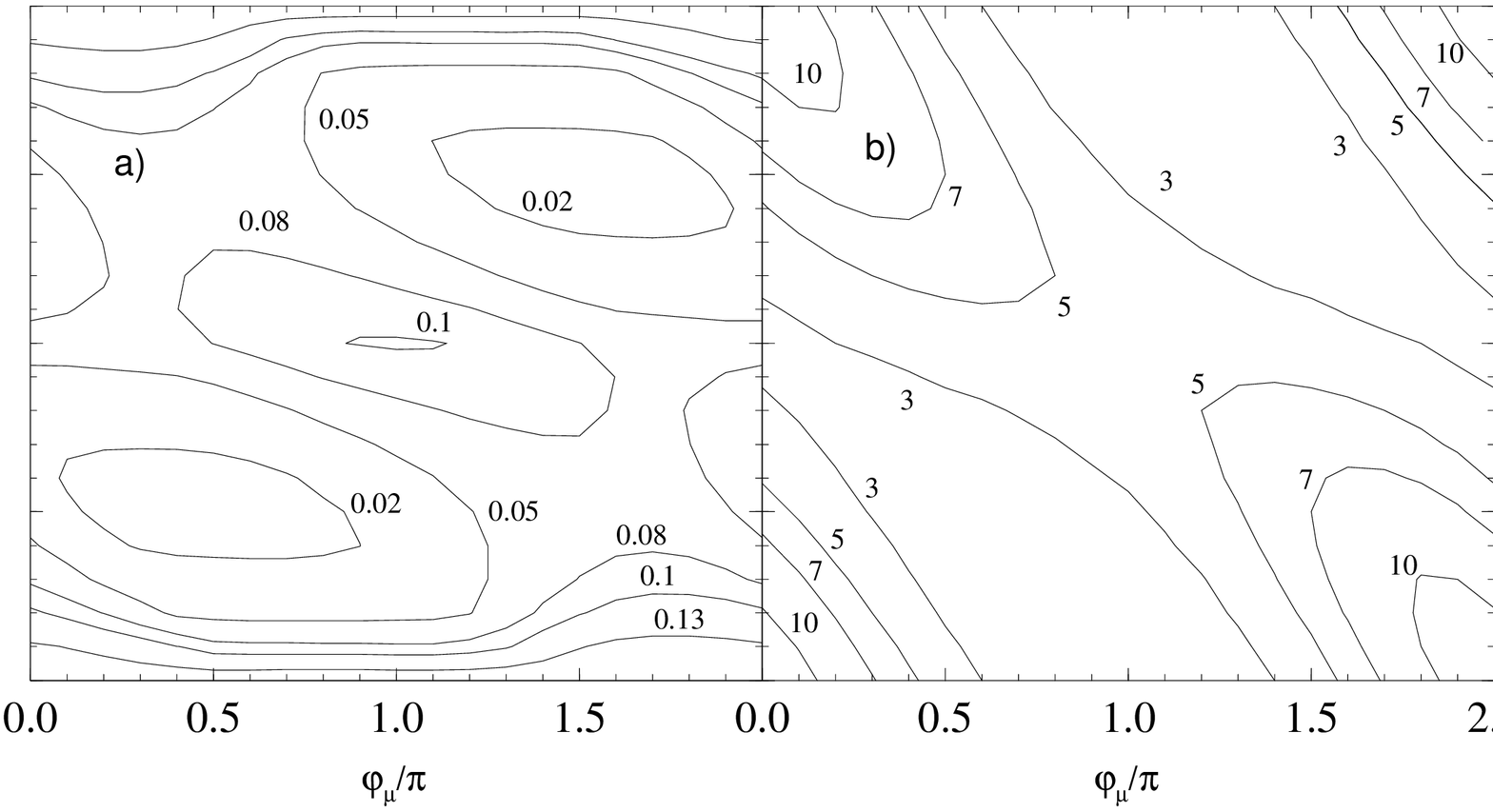}
\end{center}
\vspace{-.25in}
\hspace{2.5in}Figure 11.2


There are also other difficulties in interpreting any direct detection
results, such as relating cross sections on nuclei to cross sections on
protons and then quarks, figuring out the local galactic density of WIMPs and their
velocity distributions, and so on.  The issues raised here remain even
if all these other questions are answered.
Figure 11.1 $\cite{D. J. Chung}$ shows the region in the $\Omega _{LSP}h^{2}-\sigma _{p}$ plane
where results could occur as various soft parameters are varied; $\sigma
_{p} $ is the LSP-proton scattering cross section, a useful measure of the
strength of LSP interaction.  A reference point is shown as a dark
square.  The allowed region if soft parameters are measured to 5\% is
grey, and the allowed region for 20\% measurements is shown by the
dots.  If the soft parameters are measured to an
accuracy of about 5\% then $\Omega _{LSP}$ is determined by laboratory data
to a good accuracy (the Hubble parameter $h$ will be accurately known by the
time it is needed for this physics).  

\begin{center}
\includegraphics[scale=.65]{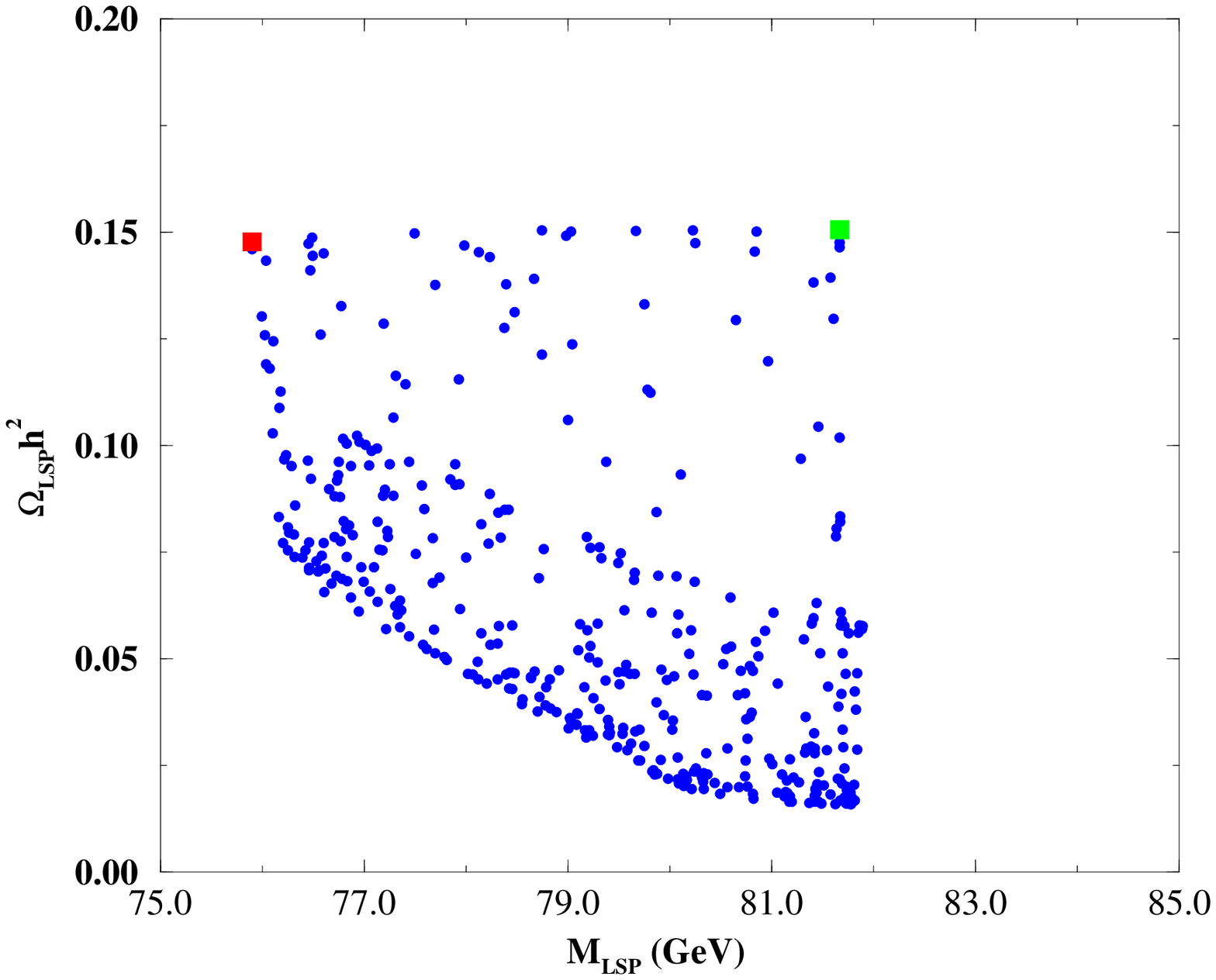}
\end{center}
\hspace{2.5in}Figure 11.3

It would be very exciting if the CDM
of the universe could be explained by particle physics laboratory data. 
This figure illustrates how direct detection (in the DAMA or CDMS or other
experiments) does not determine $\Omega _{LSP}$ --- for example, imagine a
line across the figure at a ``measured'' value of $\sigma _{p}$.  The relic
density varies by a factor of five or more along such a line.  For this
figure the soft phases were set at 0,$\pi .$

Figure 11.2 $\cite{D. J. Chung}$ shows how the relic density (the left
side) and proton-LSP cross section (the right side, in units of
$10^{-6}$ pb) vary with the phases relative to one reference point.  This can introduce an additional factor of seven or
so in the value of $\Omega _{LSP}$, so that the phases also must be measured
to 5-10\% accuracy before we can know the relic density even after the LSP
is observed.

Figure 11.3 $\cite{D. J. Chung}$ shows a scatter plot of the variation of $\Omega _{LSP}$ and the
LSP mass as the phases are allowed to vary.  The LSP mass is not very
sensitive to the phases for the soft parameters used here.  But imagine a
vertical line at a given value of the LSP mass along which the relic density varies
by a factor of about seven (remember that errors on the LSP mass will mean
including the whole range shown).  So measuring both the LSP mass and the
LSP-proton cross section will still allow a factor of 5-10 variation in the
relic density.  Only measuring the soft parameters themselves, to a good
accuracy, allows a meaningful determination of the relic density. As we will discuss
below, the needed measurements of the soft parameters including the phases
can only be done with data from a lepton collider with a polarized beam, and
energy above the threshold for producing lighter charginos and neutralinos.

\section{A STRING-MOTIVATED \ (D-BRANE)\  MODEL FOR THE SOFT PARAMETERS}

Now that we have looked at a number of examples, we want to return
to the general structure of ${\cal L}_{soft}$.  Although most of our
considerations are quite general, it is always very important to have an
explicit model in hand to test ideas and results.  It is also important to
have a model to illustrate how the many parameters of ${\cal L}_{soft}$
reduce to a few once there is theoretical structure.  We also want a string
based model since our ultimate goal is connecting the measured ${\cal L}_{soft}$ to string theory.  We want a model where the phases are
not forced to be zero, and gaugino masses are not forced to be degenerate. 
We want a realistic embedding of the SM so we can include EWSB, and no
exotic particles already excluded by experiment. At the present time no
models based on specific compactifications are satisfactory. \ The approach
of Ibanez and collaborators $\cite{See A. Brignole,L. Ibanez}$ is from our point of view a good way to
proceed. They introduce a fruitful parameterization of supersymmetry
breaking (Eq. 12.1, 12.2), allowing the calculation of ${\cal
L}_{soft}$ in particular string models.   We $\cite{Lisa Everett}$ specifically use
the results of ref. 29,30.

Here we'll focus on orbifold compactifications of Type I string theory
(which can be viewed as orientifold compactifications of Type IIB string
theory).  In such models with open (type I) string sectors, consistency
conditions require the addition of D-branes, on which the open strings
end.  The number and type of D-branes depends on the chosen
orientifold.  In this approach it is assumed that supersymmetry is
broken on other branes, and the breaking is transmitted to the SM branes
by F-term vev's of closed string sector fields (the dilation $\it S$ and
moduli $\it{T_i}$); the form of the F-term vev's of $\it S$ and
$\it{T_i}$ is specified below.  We can achieve our goals by using two
intersecting 5-branes, as illustrated below:

\begin{center}
\includegraphics[scale=.7]{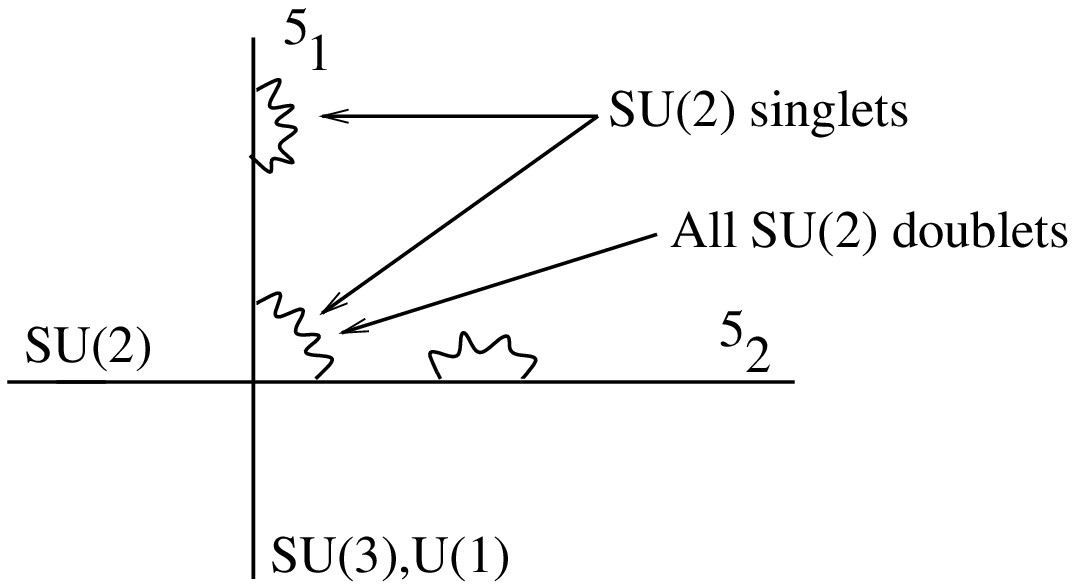}
\end{center}

The strings representing the SM particles are massless modes and thus of
zero length, but are shown for illustrative purposes in this figure. 
 We choose to embed the SM gauge group such that SU(3) and U(1) are
associated with one of the intersecting branes, and the SU(2) is from
the other.  Thus all the doublets, quark and lepton and Higgs, must touch both
branes at the intersection since they have both SU(2) and U(1) quantum numbers. 
The SU(2) singlets can be on either brane or possibly connecting them at the
intersection.  Given the basic approach and our goals as stated above
 this embedding is
essentially unique.  Other brane alternatives with the same embedding
of the gauge groups are related to this by
redefinitions and dualities.  Following references 28 and 29, the F-term
vev's are taken to be

\begin{eqnarray}
\hspace{.15in}F^{S}=\sqrt{3}(S+S^{\ast})m_{3/2}X_{0}e^{i\alpha _{S}}
\end{eqnarray}
\begin{eqnarray}
\hspace{.15in}F^{T_{i}}=\sqrt{3}(T_{i}+T_{i}^{\ast })m_{3/2}X_{i}e^{i\alpha _{T_{i}}}.
\end{eqnarray}

The $X_{i}$ satisfy $\sum_{i=0}^{3}X_{i}^{2}=1.$ Physically, the $X_{i}$
measure the relative contributions of the dilaton and moduli to the
supersymmetry breaking.  Given these F-term vev's one can calculate soft
terms with the knowledge of the superpotential, Kahler potential, and
gauge kinetic function $f_\alpha$ (in which $\alpha$ labels the gauge groups).

To illustrate how the soft terms arise it is worthwhile to write down two
results.  First, we express the superpotential and Kahler potential for
the matter fields $\Phi$ in the theory in a general (string-motivated) form as
$W=Y_{\alpha \beta \gamma}\Phi_{\alpha}\Phi_{\beta}\Phi_{\gamma}$,
$K=K_0+\tilde{K}_{\alpha}|\Phi_{\alpha}|^2$  Given the parameterization
of the F-term vev's above, the soft terms can be calculated using
(standard) super-gravity techniques.  Writing

\begin{eqnarray}
-{\cal L}_{soft}=\tilde{A}_{\alpha \beta\gamma }\Phi _{\alpha
}\Phi_{\beta }
\Phi _{\gamma }\;+\;\frac{1}{2}M_{a}\lambda _{a}\lambda _{a}\;\;........
\end{eqnarray}

the results for the soft triscalar couplings $\tilde{A}$ and gaugino
masses $M_a$ are

$$\tilde{A}_{\alpha \beta \gamma}=-{\sum_{i=0}^{3}}
\left(\frac {\partial^{2}K_0}{\partial T_{i}\partial T_{i}^{\ast}}\right)^{-1/2}\left
[\sum_{l=\alpha
\beta \gamma}\frac{\partial \ln \tilde{K}_l}{\partial
T_i}-\frac{\partial}{\partial T_i}-\frac {\partial K_0}{\partial T_i}
\right]\times$$
\begin{eqnarray}
\hspace{.15in} Y_{\alpha \beta \gamma}\sqrt{3}m_{3/2}X_{i}e^{-i\alpha_i}
\end{eqnarray}

\begin{eqnarray}
M_{\alpha}=\frac{\sqrt{3}m_{3/2}}{{2}{Re}f_{\alpha}} 
\left [\sum_{i=0,3}\frac{\partial f_{\alpha}}
{\partial T_{i}}\left(\frac{\partial^{2}K_{0}}{\partial T_{i}\partial T_{i}^{\ast
}}\right)^{-1/2}X_{i}e^{-i\alpha _{i}}\right] 
\end{eqnarray}

These results illustrate how the phases enter when the F-term vev's are
complex, and also how if the Yukawa couplings are complex then it is
natural for the soft trilinear couplings to be complex as well.  The
Yukawa couplings can in principle depend on the moduli.  Here the
$\tilde{A}_{\alpha \beta \gamma}$ are labeled by the scalar fields;
these have to be rearranged into the $3\times 3$ flavor matrix to
compare with the usual form for the triscalar couplings.

Equations 12.4, 12.5 will not be used in the following, and we will not go
into detail about their derivation.  They are shown only to illustrate
for the reader how Yukawa's and their phases, and the F-term vev's, can
enter into determining the soft terms.

The results $\cite{L. Ibanez,Lisa Everett,Gordy}$, see also 
$\cite{S. Khalil}$, for the soft parameters in our example are

\begin{eqnarray}
M_{1}=M_{3}=\sqrt{3}m_{3/2}X_{1}e^{-i\alpha _{1}}\\
M_{2}=\sqrt{3}m_{3/2}X_{2}e^{-i\alpha _{2}}\\
m_{Q_{a}}^{2}=m_{L_{a}}^{2}=m_{H_{u,d}}^{2}=m_{3/2}^{2}
 (1-\frac{3}{2}(X_{0}^{2}+X_{3}^{2}))\\
m_{U_{1}}^{2}=m_{D_{1}}^{2}=m_{E_{1}}^{2}=m_{3/2}^{2}
 (1-3X_{0}^{2})\\
m_{U_{2}}^{2}=m_{D_{2}}^{2}=m_{E_{2}}^{2}=m_{3/2}^{2}
 (1-3X_{3}^{2})\\
m_{U_{3}}^{2}=m_{D_{3}}^{2}=m_{E_{3}}^{2}=m_{3/2}^{2}
 (1-3X_{2}^{2}).
\end{eqnarray}

In general, the soft trilinear couplings have a hierarchical structure
related to that of the Yukawa couplings, but are not directly
proportional to them.  In our case, the general form of the $\tilde{A}_{a,b}^{u,d,e}$ is

\begin{eqnarray}
\hspace{.15in}\tilde{A}
^{u,d,e}\sim \left( 
\begin{array}{ccc}
0 & 0 & A_{u,d,e}Y_{13} \\ 
0 & 0 & A_{u,d,e}Y_{23} \\ 
0 & 0 & A_{u,d,e}Y_{33}
\end{array}
\right) 
\end{eqnarray}

with the model giving $-A_{u,d,e}=M_{1}=M_{3}$.  Note that the smaller
elements of $\tilde{A}_{u,d,e}$ depend on how the smaller elements of
the Yukawa couplings are generated.

If instead of two 5-branes we had used a 9-brane and a 5-brane we would have
a T-dual picture with all physical results the same, with the substitution
of $X_{0}\rightarrow X_{3},$etc.  (see for example the discussion in
$\cite{L. Ibanez}$).

The mechanism which provides the effective $\mu$ parameter in these
models is not yet known definitively but is likely to be
model-dependent, so for now it is better to simply parameterize it with
a magnitude $\mu$ and phase $\phi_\mu$.  At this stage the model
depends on 8 parameters,

\begin{eqnarray}
m_{3/2},\;\alpha_{2}-\alpha_{1},\;X_{1},\;X_{2},\;X_{3},\;\mu,\;
\phi_{\mu},\;{\rm and} \tan \beta.
\end{eqnarray}

With this model, one can do all the usual collider and cold dark matter
phenomenology, and also address aspects of CP violation physics and some flavor physics. 
The 105 parameters of the soft Lagrangian has collapsed down to this
workable number.  Once there is data the same kind of result is expected.

With our embedding gauge coupling unification is not automatic because the SM gauge
groups are not all embedded on the same brane.  If the real parts of the
moduli vev's are all about the same size one will find approximate gauge
coupling unification.  Given the long-standing 10-15\% discrepancy between
the measured value of $\alpha _{3}$ and the value in the MSSM from naive
running of the gauge couplings, this may actually be an advantage of this
model, and may usefully constrain the parameters.

These kinds of models are attractive new ways to build less naive
models that incorporate considerably more physics, both in the foundations
and phenomenologically.

\section{WHERE ARE THE SUPERPARTNERS AND HIGGS BOSON --- FINE TUNING?}

Superpartners get mass from both the Higgs mechanism and supersymmetry
breaking, the latter entering through the soft masses.  For example, in the
chargino mass matrix the off-diagonal elements come from the EW symmetry
breaking when the Higgs fields get vev's, and the diagonal elements come
from ${\cal L}_{soft}$.  The EW contributions are typically of order $M_{W}$
or less.  If the soft masses are large, the superpartner masses will
generally be large.

There are four arguments that have been used to suggest that some
superpartners will be light, i.e. near the EW scale.  The strongest
argument is from the success of supersymmetry in explaining the
electroweak symmetry breaking.  The EW symmetry breaking allows us to express 
$M_{Z}$ in terms of soft masses, so it gives one
relation $\cite{J. Ellis,G. Giudice,D. Wright,Kane}$ of the soft masses
to a known mass.  That alone is not enough to fix the soft masses, since
several terms enter the relation (as we will discuss below), so one must
impose some reasonable condition about cancellations among
the contributions.  There are no places in physics where large fine-tuning
occurs or is acceptable once there is a theory, so it is appropriate to
impose such a condition.  What appears as fine-tuning is theory-dependent.
 The usual example is the precise equality of the electric charges of
proton and electron, so atoms are neutral to a part in about 10$^{20}$.  If
electric charge is quantized that is reasonable, if not it requires a huge
fine-tuning.  So we expect any acceptable theory to imply quantization of
electric charge.  Similarly, when we judge fine-tuning of soft masses we
should do that in the presence of a theory that can relate the parameters. 
More generally, large cancellations only arise when they are exact, as
for electric charge.  For generic quantities in a theory there can be
significant cancellations, but not very large ones.  Even then, because parameters have different physical origins, and run
differently from the unification scale to the weak scale, constraints
remain.  \emph{If }supersymmetry is indeed the explanation for EWSB, then
it is appropriate to impose reasonable fine-tuning constraints on the soft
parameters.  We will see that this implies gauginos and the Higgs boson
should be light.

People have also argued $\cite{See L. Roszkowski}$ that some superpartners, most likely sleptons,
should be light or the LSP would annihilate too poorly and the large number
of LSPs left would overclose the universe.  But this assumes the LSP is the
CDM --- while I agree with that, it is an extra assumption.  Further, there
can be loopholes $\cite{M. Dress}$ from annihilation through a resonance or along particular
directions in parameter space.  A third argument is that if the baryon
asymmetry is all or part from the EW phase transition it is
necessary $\cite{M. Car,Some recent articles}$ for
charginos and stops to be lighter than about $m_{top}$, and for Higgs bosons
to be fairly light.  Again, that seems likely, but requires assuming the
baryon asymmetry is indeed produced this way.  Finally, light stops occur
in most models because the stop mass matrix tends to have four elements of
about the same size and therefore a small eigenvalue.  While the reasons of this paragraph may indeed be
relevant, the biggest success (as we discussed above) of supersymmetry is
its explanation of the mechanism of the breaking of the electroweak
symmetry, and we should be able to rely on the implications of that result.

The simplest way to write the resulting relation $\cite{Kane}$ between
$M_{Z}$ and the soft parameters is, for the SSM,
\begin{eqnarray}
M_{Z}^{2}\approx 7.2M_{3}^{2}-0.24M_{2}^{2}+0.014
M_{1}^{2}-1.8\mu^{2}-1.4M_{H_{U}}^{2}+m_{Q}^{2}+m_{U}^{2}+...
\end{eqnarray}

Here the soft parameters on the RHS are values at the unification scale, and
the coefficients are determined by RGE running to apply at the weak scale. 
These numbers are for $\tan \beta =2.5$; for other values of $\tan \beta $
the coefficients change a little.  The gluino mass is about $2.9M_{3}$. 
Note that there is significant sensitivity to the gluino mass, and $\mu $,
and less sensitivity to $M_{2}$ and $M_{1}$ or to the scalar masses. Details of the RGE running imply there is even less sensitivity to the
scalar masses than is suggested by the coefficients here $\cite{recent study}$.  Clearly, if all
of the parameters on the RHS are independent then several of them cannot be
too large or supersymmetry is not actually explaining EWSB.  But as we have
discussed, in a real theory we expect relations among the soft parameters. 
Exact cancellations are not expected, since the parameters run differently
down from the unification scale, and the coefficients depend on $\tan \beta $
.  Also, the physical origin of $\mu $ is likely to be different from the
origin of the other soft masses so precise cancellations involving $\mu $
are unlikely. 





The sensitivity to $M_{3}$ implies $\cite{Kane}$ the gluino cannot be too heavy if
supersymmetry explains EWSB.  Note that if one assumed degeneracy of the
gaugino masses $M_{i}$ then one would also assume charginos and neutralinos
must be very light, but that is only true if there is strict degeneracy. 
The chargino and neutralino masses are still constrained to be fairly light
because they also depend on $\mu ,$ as the figures below show.
We $\cite{Kane}$ have
studied how heavy the gaugino and Higgs masses can be if supersymmetry does
explain the EWSB even in a theory with cancellations, by using the D-brane
model of section 12 which allows significant cancellations.  The results are shown in
Figures 13.A, 13.B.  The grey bands show the allowed regions.  The
curves of each type are for tan$\beta$ = 2,3,10.  The amount of fine
tuning is measured by the quantities $\Delta _{a}$, for a parameter $a$,
where $\Delta _{a}=\left| \frac{a}{M_{Z}^{2}}\frac{\partial M_{Z}^{2}}{
\partial a}\right| $.  The vertical axis in the figures is the\emph{
largest}\;$\Delta _{a}$.  In the bands the other soft parameters are varied.
 For these figures the phases are set to $0,\pi $ since there is not too
much effect from varying them, except at small fine tuning where the
cancellations are strong.

\begin{center}
\includegraphics[scale=.5]{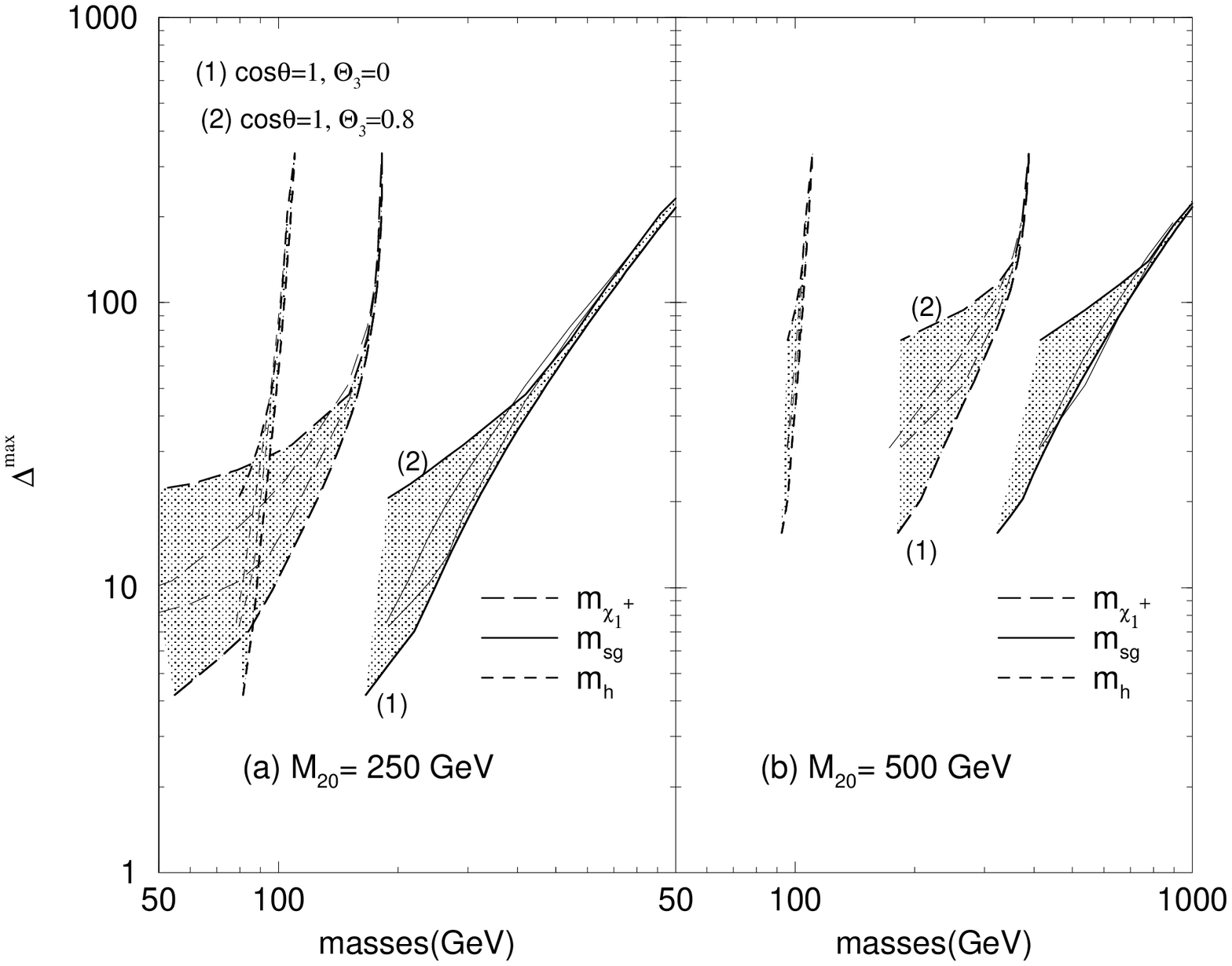}

Figure 13.A
\end{center}

\begin{center}
\includegraphics[scale=.5]{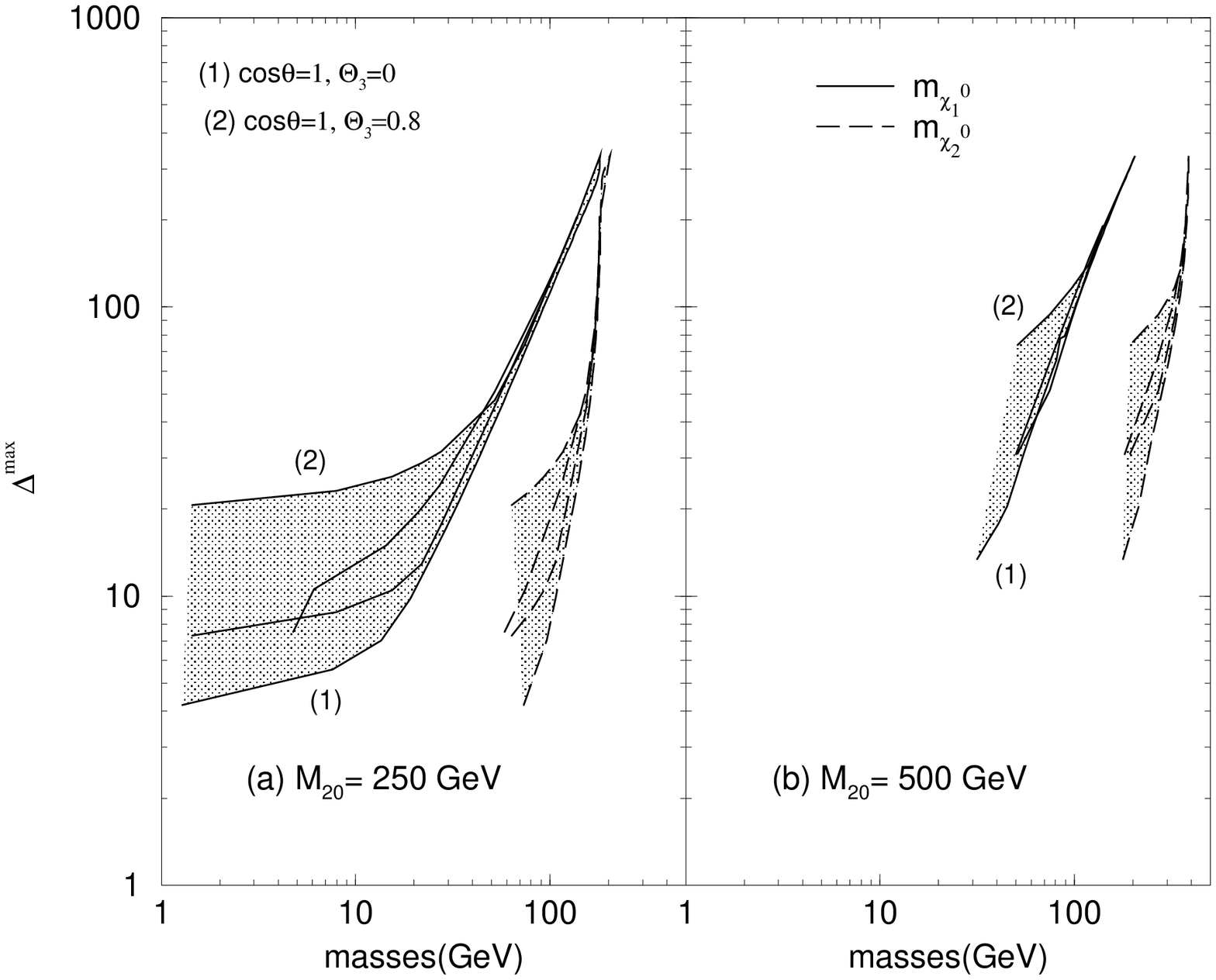}

Figure 13.B
\end{center}

Sometimes people are uncomfortable about choosing an upper limit on the
amount of fine-tuning, but if supersymmetry really is the explanation for
the EWSB then there really cannot be much fine tuning.  In that case, we can
see that gluinos cannot be heavier than about 500 GeV, which is accessible
at the Tevatron for most decay chains, the lighter chargino cannot be
heavier than about 250 GeV, again accessible at the Tevatron, the LSP cannot
be heavier than about 100 GeV and the second neutralino cannot be heavier
than about 250 GeV, so the sum of their masses is always accessible at the
Tevatron.  The Higgs boson cannot be heavier than about 105 GeV (but
remember that the current limits from LEP are only about 85 GeV when
evaluated correctly as explained in Section 7).  Note that the significant constraints on
superpartner masses apply only to gauginos and Higgs, not to squarks and
sleptons $\cite{recent study}$.

If supersymmetry really is the explanation of the electroweak symmetry
breaking then superpartners will be produced at the Tevatron.

\section{APPLICATION --- WHAT ENERGY LEPTON COLLIDER IS SAFE TO BE CONFIDENT IT CAN DO SIGNIFICANT NEW PHYSICS EVEN AFTER THE TEVATRON AND LHC?}

If there were no limit to available resources we would not have to discuss
questions such as this.  A multi-TeV lepton collider with a polarized beam
could clearly do important new physics even after LHC.  But such a collider
may not yet be technically feasible, and is certainly too expensive to build
for many years.  On the other hand, it is known how to build a 500-600 GeV $
e^{+}e^{-}$ collider now, and (although there is no official cost estimates
for such a collider) experts I have confidence in tell me that such a
collider could be proposed and built beginning now, for a cost that appears
to be reasonable given funding realities.  But is such a collider, at
most 3 times the LEP energy and considerably less than the LHC energy, worth
building for the physics results it can provide after the Tevatron and LHC
have taken data?  In the best case such a collider could be taking data in
2010.  The answer is yes, given reasonable guarantees (which have not yet
been made precise) about luminosity and polarization.  That is, a 600 GeV
linear electron collider is as safe a bet as one can make in physics for a
facility that will provide major new physics even after LHC $\cite{See my
talks}$.  Let us
call this collider the  N(ext)P(olarized)L(epton)C(ollider).  The energy
600 GeV emerges as a safe upper limit independently from both the Higgs
sector and the SUSY sector, so it would be good if an initial (say)
500 GeV collider were extendable to about 600 GeV.

The reasons why such an energy is safe comes in two parts.  One involves
the Higgs sector.  The other involves supersymmetry, and depends
strongly on understanding the soft-breaking Lagrangian, so it is a
relevant application for these lectures.

First consider the Higgs sector argument briefly $\cite{James D. Wells}$
There is now a well-established upper limit on $M_{h}$\ 
 in the SM, from precision measurements, $M_{h}\alt 200$ GeV $\cite{LEP}$.
  If a light Higgs boson in fact exists the bound is satisfied and NPLC can
study it, and make a number of significant measurements that cannot be made
at the Tevatron or LHC, such as the branching ratios to up-type quarks and
to gluons.  Even more importantly, if there is not a light Higgs boson but
other physics that also contributes to the precision data in just the right
way to fake the upper limit, then the tight limit plus good understanding of
the theory allows $\cite{James D. Wells}$ one to be sure that (a) there is a scalar state that can
be studied at NPLC with a mass of at most 500 GeV (500 is an upper limit --
usually its mass is less.  So to find this scalar probably 500 GeV
is enough, but 600 GeV is safe.  Further, (b) there must always be an
associated piece of physics that affects the precision data in just such a
way as to compensate for the effects of the heavier scalar.  This
additional contribution, a heavy fermion or Kaluza-Klein states, or
something else, will also have effects that are detectable at NPLC (that may
require running NPLC at the Z mass, but such running is expected at any
linear electron collider if it is scientifically justified).  Thus in the
Higgs sector NPLC600 is a very good facility to have if there is indeed a light
Higgs boson as is expected, and it is even better if there is no light
Higgs.  This argument provides strong justification for such a collider.

Next turn to the physics justification based on supersymmetry.  We have
 seen in section 13
that if supersymmetry is indeed the explanation for the fundamental question
of how the electroweak symmetry is broken so that particles can have mass in
the Standard Model in a consistent way, then some superpartners (in
particular, charginos and neutralinos) are not too heavy.  In fact, at
least the lightest chargino and the lighter two neutralinos could be studied
at NPLC600.  (Perhaps the light stop and sleptons could also be studied
at NPLC600.)  This is extremely important because it allows the measurement of $%
\tan \beta $ and the soft phases, which could not be done at LHC.

To clarify this very important point, let us elaborate on what NPLC600 would
be able to achieve that hadron colliders could not. There are four such
physics opportunities in the supersymmetry sector. The first three are important, and the fourth
is by far the most important.  First, if nature is not supersymmetric at
the weak scale it is essential to know that so we can focus on other
directions.  Knowing that would have a major effect on string theory
activities as well. If signals for supersymmetry are not found at the
Tevatron and LHC it will be hard to still think that low energy
supersymmetry is part of nature, but because of the large backgrounds it is
possible to hide supersymmetry signals at hadron colliders.  Only a
lepton collider above the threshold for some superpartners can be definitive
here. Why should 600 GeV be above that threshold?  To repeat, if
supersymmetry does explain the EW symmetry breaking we saw above that
600 GeV should be a safe upper limit.  If supersymmetry does not explain
the EWSB it is unlikely to be part of the low energy description of
nature.  

Second, it is important to be able to demonstrate $\cite{A number of such}$ that a set of signals is
indeed supersymmetry.  The spins of the superpartner candidates need to be
measured.  Many relations among coupling strengths predicted by
supersymmetry need to be tested.  For example, the quadratic Higgs coupling 
$\lambda $ should be given by gauge couplings $\left(
g_{1}^{2}+g_{2}^{2}\right) /2$, the $\tilde{g}\tilde{g}g$ coupling should be
the same as the gluon self coupling $g_{3}$, the $\tilde{d}d\tilde{\gamma}$
coupling should be $\frac{1}{3}e$, the neutralinos and gluino should be
Majorana particles, and so on.  It is extremely hard to demonstrate any of
these results at a hadron collider, while it is doable at NPLC600.  Third,
some superpartners or some BR of some superpartners or Higgs bosons may be
invisible at hadron colliders, but detectable at NPLC600.  That can happen
if masses are close to degenerate, if some decays are invisible such as $%
h\rightarrow LSP+LSP\;($or $h\rightarrow $Kaluza Klein modes or some other
exotic channel), $\tilde{N}\rightarrow \tilde{\nu}\nu $, and so on. 
Various chargino, neutralino, Higgs, and stop modes can be in this category.

Finally we consider the most important point.  Let us assume that
superpartners are indeed observed.  As we have seen in these lectures, the central problem of particle physics
is then to measure and understand the soft-breaking supersymmetry Lagrangian
(and $\tan \beta $), and to use it to help formulate and test string theory.
We have seen in these lectures that it is essential to learn the patterns
of ${\cal L}_{soft}$, whether the soft masses are degenerate, whether some
or all phases are large or not, and also the numerical values of the soft
parameters.  These quantities are at least as fundamental as the quark masses and
the CKM mixing angles and phase.  

Learning the quark masses and mixing angles did not lead to a breakthrough
in understanding.  Most likely that is because at the level of the string
theory some Yukawa couplings are of order the gauge couplings and others are
zero.  Then those that are zero get small non-zero values as some
symmetries are broken, and/or from non-perturbative operators and loop
corrections.  It may be that we have to understand a great deal of the
theory before they can be explained.  It is likely that the situation with
the soft parameters will be simpler since all the superpartners get mass
from them, so the leading effects will be important, and they will tell us
explicitly about how supersymmetry  is broken, and about compactification.

To measure the soft parameters it is necessary to have enough
observables.
 Think about hadron colliders.  There is much that hadron colliders will
do well, such as measure the dominant quantum numbers of the LSP and
therefore largely determined how supersymmetry breaking is mediated, and
measure the mass of the LSP $\cite{S. Mrenna}$ which is relevant for
learning whether the LSP is the cold dark matter.  One can measure the
 masses of most (though probably not all) of the
superpartners.  We have seen that at most the
gluino (approximately) and gravitino (if it is the LSP) masses are directly
connected to those in the Lagrangian; for the rest one must invert equations
relating the Lagrangian parameters to the masses and other soft parameters
including the phases and $\tan \beta$, as in the chargino example of
Section 5 or the Higgs example of Section 7.  One can measure $\sigma \times BR$ for
most superpartners, though certainly not all since the backgrounds are large
and many signatures overlap.  Since there are 33 superpartner and Higgs
states, this could give 66 observables, though that is surely an
overestimate, considerably fewer than the number of parameters.  If one could measure differential cross sections in bins
with significant statistics one would of course have more observables.  But
in practice that is probably not possible --- even at electron colliders
usually only forward-backward asymmetries are measured.  So the total
number of observables at hadron colliders is very unlikely to be enough to
invert and solve for the soft parameters and $\tan \beta .$  Possibly we
will be lucky and the masses and BR will have values that allow some limited
sector to be solved for its soft parameters, but we cannot count on
that.  (There has been some work reported on learning supersymmetry
parameters at LHC, but always in the context of models assumed to be
known, and with a number of assumptions about values of most
parameters.  Once there is data, of course experimenters will want to
measure parameters without such assumptions.)

Lepton colliders are often characterized as being cleaner than hadron
colliders, and that is true, but that is not the main reason they are
important for the future of particle physics.  The main reason is that they
allow the measurement of more observables.  The polarized beam immediately
doubles the number of observables.  Equally important, running at a second
energy again doubles the number of observables, so effectively there are
four times as many observables at a polarized lepton collider as at a hadron
collider!  Of course that is only true for the superpartners that are
produced at the lepton collider, so some observables are lost because the
relevant masses are too large.  To get the full picture it is essential to
have both the lepton collider and the hadron colliders.  In fact, both LHC
and the Tevatron are essential because the Tevatron produces the lighter
superpartners and can tell us about their masses and BR, allowing us to then
untangle the complicated situation at the LHC where many superpartners will
be produced and the heavier ones will have complicated decay chains through
several intermediate steps.  The Tevatron is complimentary to both LHC and
NPLC600, to the latter in part because it produces colored states.  Of
course, finally results from low energy experiments such as CP violation at
b-factories, EDMs, rare decays, proton decay, cold dark matter and more will
be combined with collider data to fully learn the soft Lagrangian.

There are no realistic studies of how many observables NPLC600 can measure.
 Simulations under conditions where no favorable assumptions about decay BR
are made, and with reasonable experimental errors and statistics included,
are essential to confirm the qualitative analysis above.  But given the
existence of the Tevatron and LHC and b-factories and serious efforts on the
other low energy experiments, it is likely that NPLC600 is both necessary
and sufficient to let us measure enough about the world to proceed to
formulate and test string theory.

\section{ACKNOWLEDGEMENTS}

I am grateful to M.Brhlik, D.Chung, K.Dienes, S.King, S.Mrenna,
L.-T.Wang, and J.Wells for discussions and suggestions, and I
particularly appreciate discussions, suggestions and contributions to
these lectures from L.Everett.

\end{document}